\documentclass[10pt]{article}

\usepackage[utf8]{inputenc}
\usepackage{microtype}
\usepackage[inline]{enumitem}
\usepackage{pifont}
\usepackage[accepted]{tmlr}

\usepackage{amsmath,amsfonts,bm}

\def\eqref#1{equation~\ref{#1}}

\def\1{\bm{1}}

\DeclareMathAlphabet{\mathsfit}{\encodingdefault}{\sfdefault}{m}{sl}
\SetMathAlphabet{\mathsfit}{bold}{\encodingdefault}{\sfdefault}{bx}{n}

\usepackage{hyperref}
\usepackage{url}
\usepackage{inconsolata}

\usepackage{graphicx}
\usepackage{amsmath}
\usepackage{amssymb}
\usepackage{array}
\usepackage{booktabs}
\usepackage{colortbl}

\definecolor{panelshade}{RGB}{246,235,232}
\definecolor{tablerow}{RGB}{245,245,245}

\title{Towards Realistic Evaluation of Uncertainty Quantification for LLM-Based Recommender Systems}

\author{\name Hao Wu \email hwu522@wisc.edu \\
      \addr University of Wisconsin--Madison
      \AND
      \name Sipeng Chen \email sipengchen2028@u.northwestern.edu \\
      \addr Northwestern University
      \AND
      \name Weikai Li \email weikaili@cs.ucla.edu \\
      \addr University of California, Los Angeles
      \AND
      \name Tianyi Chen \email tchen523@wisc.edu \\
      \addr University of Wisconsin--Madison
      \AND
      \name Qiancheng Yang \email qyang333@wisc.edu \\
      \addr University of Wisconsin--Madison
      \AND
      \name Hang Zhou \email hzh@unc.edu \\
      \addr Department of Statistics and Operations Research \& School of Data and Information Sciences  \\
      The University of North Carolina at Chapel Hill
      \AND
      \name Yizhou Sun \email yzsun@cs.ucla.edu\\
      \addr University of California, Los Angeles
      \AND
      \name Wei Wang \email weiwang@cs.ucla.edu \\
      \addr University of California, Los Angeles
      \AND
      \name Xiao Luo \email xiao.luo@wisc.edu \\
      \addr University of Wisconsin--Madison
}

\def\month{11}
\def\year{2026}
\def\openreview{\url{https://openreview.net/forum?id=GXDe6REdu3}}

\begin{document}
\maketitle
\begin{abstract}
Large language models (LLMs) provide a flexible interface for recommendation systems. However, their issued rankings require reliable instance-level uncertainty monitoring. In this scenario, comparing uncertainty quantification (UQ) methods becomes challenging because LLM rankers lack a standard item-classification head. Their scores may depend on option labels, prompts, generation procedures, and the available user and item evidence. Existing methods derive uncertainty from different sources and apply it at different stages of the recommendation process. Some methods evaluate an issued ranking, whereas others use uncertainty to change
the ranking itself, making direct comparison difficult. Here we introduce a fixed-ranking benchmark which evaluates eight heterogeneous UQ signals as record-level monitors under a shared output contract. We evaluate these signals across four recommendation datasets and two LLM backbones under a shared fixed-ranking protocol. From the results, we draw three main observations. First, UQ method ordering changes across backbones even on the same reference instances. Second, ranking-aligned Top10Mass provides the strongest scalar Error@10 detection in several evaluation conditions, reflecting effective cutoff-aligned monitoring, while its advantage narrows for popularity-tail targets. Third, scalar UQ signals
are weak at identifying which control has the larger clipped NDCG@10 difference, whereas source-specific signals, including Perturbation Rank Instability and simple listwise distances, provide stronger paired-control agreement, reflecting their sensitivity to the two controlled evidence sources. These findings motivate us to evaluate ranking-failure detection, robustness across recommendation
conditions, and responses to controlled evidence changes as distinct properties of UQ. Code and evaluation artifacts are available in our repository:
\url{https://github.com/haowu0916edisonwu/recuq-benchmark.git}.
\end{abstract}

\section{Introduction}
\label{sec:introduction}

Large language models (LLMs) provide a flexible interface for recommendation systems by representing user histories, candidate items, and task instructions
in natural language and asking an instruction-tuned model to rank the candidates \citep{hou2024llmrank}. Such an interface can
support zero-shot or few-shot recommendation without requiring an additional recommendation-specific
prediction head for each domain. In this work, we focus on the LLM-as-reranker setting and use released instruction-tuned backbones directly, without training from scratch or recommendation-specific fine-tuning. However, a fluent and well-formed ranked list may still be unreliable. The relevant item may be missing from the top results
even when the model shows strong confidence. Recent work has begun to consider whether average ranking utility alone is sufficient for evaluating LLM-based recommendation systems \citep{jiang2025beyond}. This broader reliability perspective examines whether the model outputs remain trustworthy when the confidence is unreliable, fairness concerns arise, or recommendations are sensitive to preference and presentation biases \citep{sah2026uncertaintyfairness,tang2026biasrecbench}. Practical deployment therefore requires signals that can monitor the reliability of each issued
ranking. 

However, applying uncertainty quantification (UQ) to LLM-based ranking is not
straightforward. Unlike fixed-class prediction, recommendation produces a
user-dependent ranked list whose top-$k$ outcome is the object being monitored.
An LLM ranker may output candidate labels or the entire ranked list, and its
uncertainty can also be affected by the prompt, the option-label format, the
decoding or sampling procedure, and the user or item evidence provided in the
context. Consequently, different UQ methods may reflect different aspects of the model behavior, which makes their scores difficult to interpret and compare under a common standard. Several kinds of work have approached the uncertainty from different angles. Classical methods focus on the confidence attached to a prediction, either by using predictive probabilities to detect likely errors \citep{hendrycks2017baseline} or by examining whether reported confidence matches the empirical correctness \citep{guo2017calibration}. Language-model research extends this idea beyond a single probability score by studying the self-assessed correctness \citep{kadavath2022language} and consistency across sampled responses \citep{wang2023selfconsistency,kuhn2023semantic}. Moreover, in recommendation, uncertainty has taken on a broader role: it has been used to estimate the recommendation quality \citep{li2025selfaware}, distinguish recommendation uncertainty from uncertainty of prompts \citep{kweon2025uncertainty}, and support reranking or model optimization \citep{xu2026guider,fan2026ugr}. These methods are difficult to compare directly because they differ both in
the uncertainty signals they use and in how uncertainty enters the
recommendation process \citep{xu2026guider,fan2026ugr}. In particular, methods that use uncertainty for
reranking or model optimization also change the recommendation output, making it difficult to separate UQ monitoring quality from changes in the underlying ranking. To isolate monitoring quality, we therefore hold each issued ranking fixed and map every method to the same record-level output. This fixed-ranking design allows different UQ signals to be evaluated on the same recommendation
outcome under a common standard.

Building on this fixed-ranking design, we construct a benchmark for UQ in LLM-based recommendation. Each 20-candidate base ranking is generated once and remains unchanged during UQ evaluation. Eight heterogeneous UQ signals are compared as
the record-level monitors under a shared output contract. The full Llama-3.1-8B-Instruct evaluation \citep{dubey2024llama} spans four
recommendation datasets and four evaluation settings: a full-evidence reference condition (Warm-start), user-history truncation (Sparse-history), controlled candidate item-text ablation (Item-text robustness), and a subset of Warm-start records whose positive items fall in the popularity tail (Tail-target). We also evaluate Qwen3-8B in non-thinking mode \citep{qwen3technicalreport} on the same 4,000 reference instances to examine whether UQ method ordering is preserved on a second backbone. This comparison is limited to the reference condition and does not establish broad cross-model transfer. The benchmark defines two complementary evaluation tasks. \ding{172} \textbf{Ranking-Failure Detection} evaluates whether a UQ score can flag Error@10 cases, where the target item is absent from the fixed top-10 ranking. \ding{173} \textbf{Paired-control response agreement} compares matched user-history and item-content interventions for the same base instance and tests whether a method responds more strongly to the source with the larger clipped NDCG@10 difference. Sparse-history and Item-text robustness provide the user-history and item-content controls, respectively.

\begin{figure*}[t]
\centering
\includegraphics[width=\textwidth]{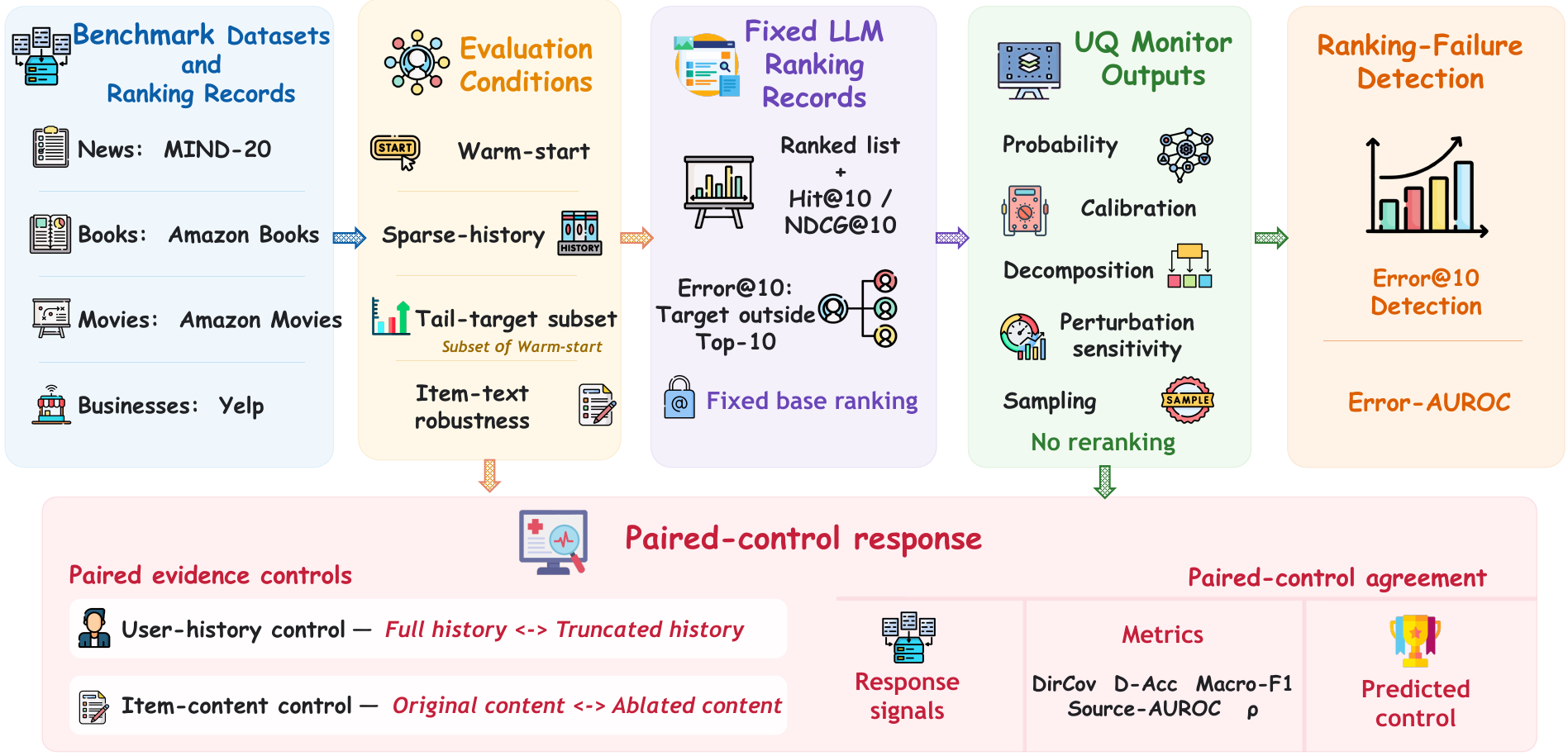}
\caption{\textbf{Benchmark pipeline.}
Fixed ranking records are scored by UQ monitors without reranking and
evaluated for Error@10 detection and scoped paired-control response agreement.}
\label{fig:benchmark-pipeline}
\vspace{-4mm}
\end{figure*}

Our experiments yield three main observations. First, the relative
effectiveness of UQ monitors is not stable across LLM backbones. Second,
ranking-aligned probability mass provides a strong scalar signal for detecting
ranking failures, but its advantage varies across evaluation conditions and
target-popularity regimes; these results support ranking-aligned monitoring for the corresponding ranking-failure objective. Third, scalar failure detection and paired evidence-response agreement capture different properties of UQ: source-specific signals are better suited to source-specific diagnosis under the clipped paired-control target, with simple non-UQ
ranking distances showing similarly strong agreement. This last finding is limited to the two constructed controls and does not establish
general source attribution. Our contributions are summarized as follows:
\begin{itemize}[leftmargin=*,itemsep=2pt,topsep=3pt]
\item{} We define fixed-ranking UQ monitoring for LLM-based recommendation
and introduce two evaluation tasks: Error@10 detection and paired-control
response agreement for user-history and item-content evidence; 

\item{} We build an extensive benchmark that compares eight UQ signals
across four datasets. It includes a full Llama evaluation suite and a Qwen
Warm-start replication on the same 4,000 instances under a shared
record-level contract; 

\item{} We provide practical guidance for evaluating UQ in LLM-based
recommendation: ranking-failure detection and paired-control response agreement
capture distinct properties of UQ, and method robustness should be assessed
across backbones and recommendation conditions because conclusions from a
single reference setting may not generalize.
\end{itemize}
\section{Related Work}
\label{sec:related-work}

\subsection{LLMs for Recommendation} 

Language-model-based recommendation builds on earlier work which formulates recommendation as a text generation task. P5 represents user interactions, item information, and different recommendation tasks in a unified text-to-text framework \citep{geng2022p5}. With the development of instruction-tuned LLMs, researchers began to examine whether a general language model could perform recommendation directly. A common setting presents the user's history, candidate items, and ranking instruction in natural language, then asks the model to return a ranked list \citep{hou2024llmrank}. Later studies adapted LLMs more closely to recommendation data and item spaces. TALLRec tunes an LLM
with recommendation instructions for the prediction of user preference \citep{bao2023tallrec}; BIGRec generates semantic item representations and maps them to actual items through a two-step grounding procedure \citep{bao2025bigrec}; GenRec directly generates target items from textual user and item information \citep{ji2024genrec}. Moreover, RecRanker tunes an LLM for pointwise, pairwise, and listwise top-$k$ ranking tasks \citep{luo2025recranker}. LLMs have also been integrated into multi-stage recommendation systems. LlamaRec first retrieves a smaller candidate set with a lightweight sequential recommender and then applies an LLM ranker to those candidates \citep{yue2023llamarec}. Together, these studies cover several ways of using language models in recommendation tasks, like task-specific tuning, item grounding, direct generation, candidate ranking, and retrieval ranking integration. More recent research has expanded the evaluation beyond average
ranking quality because strong aggregate performance does not guarantee reliable results for every user or candidate set
\citep{jiang2025beyond}. This broader reliability perspective includes uncertainty and fairness across recommendation cases
\citep{sah2026uncertaintyfairness}, as well as the sensitivity to preference and presentation biases \citep{tang2026biasrecbench}. Building on this reliability
perspective, our study evaluates whether uncertainty signals can identify unreliable issued rankings while keeping the recommender output and candidate order fixed.

\subsection{Uncertainty Quantification}

Uncertainty quantification provides signals for assessing how reliable a model’s output is \citep{gawlikowski2023survey}. Probability-based methods summarize
uncertainty through quantities such as maximum confidence, predictive entropy, and the gap between leading candidates \citep{gawlikowski2023survey,hendrycks2017baseline,settles2009active}. Calibration addresses a different question by testing whether the reported confidence matches observed frequencies of correctness \citep{guo2017calibration}. Selective prediction examines how prediction risk changes as increasingly uncertain examples are removed \citep{geifman2017selective}. These properties are related, but strong performance on one does not guarantee strong performance on the others. For language models, uncertainty can also be estimated from generated responses. Prior work studied whether a model can assess the correctness of its own answer \citep{kadavath2022language} while repeated generations reveal how stable its outputs are. Self-consistency measures agreement across sampled outputs \citep{wang2023selfconsistency}, and semantic uncertainty groups responses by meaning before measuring their variation \citep{kuhn2023semantic}. Recent studies further examine the relationship between sample consistency and confidence calibration \citep{lyu2025calibrating,xiao2025consistency}. However, these methods rely on different outputs and sampling procedures, which makes their uncertainty scores difficult to compare under a common
evaluation setup. In recommendation, uncertainty has been used for several purposes. \citet{paliwal2024predictive} study predictive uncertainty in CTR-based
recommendation through selective prediction, out-of-distribution detection, and active learning. \citet{angelopoulos2023recommendation} construct recommendation sets with finite sample of false-discovery control, and \citet{li2025selfaware} use the list-distribution uncertainty to estimate recommendation performance without relevance labels. For LLM-based
recommendation, \citet{kweon2025uncertainty} decompose uncertainty into recommendation and prompt components. Other studies use uncertainty to change the recommendation process. GUIDER adjusts the reranking decisions according to uncertainty \citep{xu2026guider}, and uncertainty-aware generative recommendation incorporates uncertainty into preference optimization and confidence learning \citep{fan2026ugr}. Unlike approaches that use uncertainty to modify the recommendation process,
our benchmark evaluates heterogeneous UQ signals after the ranking has been
issued, keeping the candidate order fixed and mapping every method to the same
record-level output contract for direct comparison.

\subsection{Robustness of Recommender Systems}

Robustness research examines whether a recommender remains reliable when its
data are sparse, imbalanced, noisy, or perturbed \citep{zhang2025robust}. Sparse user histories make it difficult to represent user preferences accurately \citep{adomavicius2005toward}. Also, popularity
imbalance may reduce the visibility and recommendation quality of tail items
\citep{abdollahpouri2019managing}. In LLM-based recommendation, these problems
appear through the textual evidence included in the prompt because missing
user history or item information can change both the resulting ranking and its
uncertainty score. Input perturbations provide a direct way to study this sensitivity.
\citet{oh2022ranklist} measure how recommendation lists change after small
perturbations to user interactions. Controlled testing provides a broader
framework for evaluating model behavior. For example, CheckList uses targeted behavioral
tests to reveal failures which the aggregate accuracy may hide
\citep{ribeiro2020checklist}. Moreover, recent LLM robustness work also examines the model
stability under controlled prompt variations which preserve the original
meaning \citep{romanou2026brittlebench}. These studies suggest that results
from one reference setting may not describe how a model behaves after its
input evidence changes. Recommendation research has also examined which inputs
contribute to uncertainty. Explainable Uncertainty Attribution measures how
individual interactions in a user's history affect predictive uncertainty in
sequential recommendation \citep{balsellsrodas2024exua}. Building on controlled robustness testing, our analysis focuses on two
predefined evidence sources, user history and item content, and evaluates
whether changes in UQ responses agree with the NDCG@10 differences associated with the two controls. This scoped evaluation differs from general
source attribution and does not claim causal identification.

\section{Benchmark and Evaluation Protocol}
\label{sec:method}

Our benchmark compares heterogeneous uncertainty quantification (UQ) signals
on fixed LLM ranking records. The base ranking is generated once and remains
unchanged throughout UQ evaluation. Figure~\ref{fig:benchmark-pipeline}
summarizes the benchmark pipeline, from evaluation conditions and fixed ranking
records to UQ monitor outputs, ranking-failure detection, and paired-control
response analysis. This section defines the problem setting, datasets and
evaluation conditions, evaluation tasks, UQ monitoring signals, and statistical
protocol. 

\subsection{Problem Definition}
\label{sec:method-problem}

Given a fixed LLM reranking record with a user history, 20 candidate items,
and a parsed base ranking, our benchmark evaluates UQ methods as record-level
monitors of the issued ranking, with the candidate order kept fixed. Each method outputs a scalar uncertainty score aligned to the same base record and may additionally provide paired-control
response scores for user-history and item-content evidence. The two controls are paired at the base-instance level: the same underlying recommendation case is evaluated under full versus truncated user history and under original versus ablated item content. Our benchmark
examines three questions: (1) whether a UQ score can identify an Error@10
case, where the positive item is absent from the fixed top-10 ranking;
(2) whether the relative performance of UQ methods remains stable across
backbones and evaluation conditions; and (3) whether the control-specific UQ responses agree with which source has the larger clipped NDCG@10 difference.

For record $i$, let
$\mathcal{C}_i=\{c_{i1},\ldots,c_{i20}\}$ denote the candidate set and
$y_i\in\mathcal{C}_i$ denote the positive item. Following LLM-as-reranker
formulations \citep{hou2024llmrank,jiang2025beyond}, the base LLM ranks the
candidate labels from a natural-language prompt. We denote the parsed ranking
by $\pi_i$ and its fixed top-10 candidate set by $B_i^{(10)}$. The ranking
$\pi_i$ remains unchanged throughout UQ evaluation, so all UQ methods monitor
the same issued ranking without modifying the candidate order.

\subsection{Datasets and Evaluation Conditions}
\label{sec:method-datasets}

We instantiate the fixed 20-candidate reranking benchmark with
Llama-3.1-8B-Instruct \citep{dubey2024llama}, following the LLM-as-reranker
setting of \citet{hou2024llmrank} and \citet{jiang2025beyond}. We additionally
evaluate Qwen3-8B non-thinking \citep{qwen3technicalreport} on the same
4,000 reference instances to examine whether method ordering reproduces on a
second backbone. Qwen is used only for this same-instance reference-condition
replication. For Amazon Books, Amazon Movies, and Yelp, we use leave-last-positive-out
evaluation, where the user's last positive interaction is treated as the
target item and evaluated against a sampled 20-item candidate set. For
MIND-20, we use the impression-based candidate sets provided by the original
benchmark, where each logged impression already contains the displayed news
candidates and the clicked item serves as the positive target. The records are constructed from MIND \citep{wu2020mind}, Amazon reviews \citep{hou2024bridging}, and the Yelp Open Dataset
\citep{yelpOpenDataset}.

Following the evaluation setting of \citet{jiang2025beyond}, we begin with
1,000 base instances per dataset. We use the original full-evidence records
as the Warm-start reference condition. Sparse-history retains the full history of up to 50 most recent positive interactions (ratings $\ge4$ for Amazon and Yelp; clicked news for MIND) and four variants keeping only the $k\in\{1,3,5,10\}$ most recent interactions, with candidates and their texts unchanged. These variants retain recent interactions and remove older interactions. Item-text robustness removes one field from all 20 candidate descriptions while leaving history unchanged: title, abstract, or category for MIND-20; title, description, categories, or author for Books; title, description, or categories for Movies; and name, categories, location, or review summary for Yelp (up to three training reviews, 240 characters each). A field variant exists only if at least one candidate has that field. These conditions supply the user-history and item-content controls. Tail-target retains Warm-start records whose targets have training-interaction counts at or below the 25th percentile over items with training interactions: MIND-20 $\le5$, Books and Movies $\le3$, and Yelp $\le2$. Targets with zero training interactions are also included. Counts use training-period clicks for MIND and positive interactions before each user's validation and test items for Amazon and Yelp.  The Llama runs produce 45,163 ranking records in total. After restricting the popularity-tail condition to the 1,312 eligible tail-target records, 42,475 records enter the final evaluation: 4,000 Warm-start, 20,000 Sparse-history, 17,163 Item-text robustness, and 1,312 tail-target records. Table~\ref{tab:benchmark-scale} summarizes the generated records by dataset.
Amazon Movies contains fewer generated variants because some item-text
ablation fields are not available for our task settings.

\begin{table*}[t]
\centering
\footnotesize
\setlength{\tabcolsep}{5pt}
\renewcommand{\arraystretch}{1.2}
\setlength{\aboverulesep}{0pt}
\setlength{\belowrulesep}{0pt}
\def\tabstrut{\rule{0pt}{1.75ex}}
\begin{tabular}{p{0.15\textwidth}p{0.11\textwidth}p{0.48\textwidth}rr}
\toprule
\rowcolor{panelshade}
\tabstrut Dataset & Domain & Base Record Form & Base & Records \\
\midrule
\rowcolor{tablerow}
\tabstrut MIND-20 & News & Clicked-news history + 20 news candidates & 1,000 & 11,000 \\
\tabstrut Amazon Books & Books & User history + 20 book candidates & 1,000 & 12,000 \\
\rowcolor{tablerow}
\tabstrut Amazon Movies & Movies & User history + 20 movie candidates & 1,000 & 10,163 \\
\tabstrut Yelp & Business & User history + 20 business candidates & 1,000 & 12,000 \\

\bottomrule
\end{tabular}
\caption{\textbf{Benchmark overview.}
Each dataset starts from 1,000 base instances; the Llama runs produce 45,163
ranking records before tail-target eligibility filtering.}
\label{tab:benchmark-scale}
\vspace{-4mm}
\end{table*}

\paragraph{Base Ranking Quality.}
\label{sec:method-base-ranking}
Table~\ref{tab:base-ranking} reports Hit@10 and NDCG@10 for the fixed Llama
rankings under each evaluation condition. These values describe the
recommendation records monitored by the UQ methods and provide the base
ranking quality before uncertainty is evaluated. The gains after removing evidence partly reflect fewer positional rankings (the candidate order or its reverse), which comprise 46.1\% of Llama Warm-start outputs. Among non-positional rankings, truncation helps mainly on Books and slightly on Yelp, consistent with recent interactions being predictive. Item-text gains are field-specific and, for Movies, partly reflect variant composition (Appendix~\ref{sec:appendix-removal}). MIND-20 and Movies have relatively weak Warm-start base rankings (Table~\ref{tab:base-ranking}). These fixed rankings determine the Error@10 records available for monitoring. Error-AUROC measures how well each signal distinguishes error and non-error records within the evaluated condition (Table~\ref{tab:uq-warm-start}). Appendix~\ref{sec:appendix-removal} reports a complementary analysis on pooled non-positional Llama Warm-start rankings.

\begin{table}[!t]
\centering
\footnotesize
\setlength{\tabcolsep}{5pt}
\renewcommand{\arraystretch}{1.2}
\setlength{\aboverulesep}{0pt}
\setlength{\belowrulesep}{0pt}
\def\tabstrut{\rule{0pt}{1.40ex}}
\begin{tabular}{lcccccccc}
\toprule
\rowcolor{panelshade}
\tabstrut & \multicolumn{2}{c}{Warm-start} & \multicolumn{2}{c}{Sparse-history} & \multicolumn{2}{c}{Tail-target subset} & \multicolumn{2}{c}{Item-text robustness} \\
\rowcolor{panelshade}
\tabstrut Dataset & Hit@10 & NDCG@10 & Hit@10 & NDCG@10 & Hit@10 & NDCG@10 & Hit@10 & NDCG@10 \\
\midrule
\rowcolor{tablerow}
\tabstrut MIND-20 & 0.546 & 0.270 & 0.557 & 0.272 & 0.557 & 0.278 & 0.539 & 0.266 \\
\tabstrut Books & 0.697 & 0.438 & 0.768 & 0.530 & 0.584 & 0.341 & 0.717 & 0.453 \\
\rowcolor{tablerow}
\tabstrut Movies & 0.579 & 0.316 & 0.594 & 0.332 & 0.566 & 0.284 & 0.614 & 0.341 \\
\tabstrut Yelp & 0.605 & 0.357 & 0.649 & 0.411 & 0.598 & 0.334 & 0.614 & 0.363 \\
\bottomrule
\end{tabular}
\caption{\textbf{Base LLM recommendation quality.}
Hit@10 and NDCG@10 for the fixed Llama rankings across datasets and
evaluation conditions.}
\label{tab:base-ranking}
\vspace{-4mm}
\end{table}

\subsection{Evaluation Tasks and Metrics}
\label{sec:method-evaluation}

\paragraph{Ranking-Failure Detection.}
For each fixed ranking record $i$, we define the Error@10 event as
\begin{equation}
    e_i
    =
    \mathbf{1}
    \left[
        y_i \notin B_i^{(10)}
    \right],
    \label{eq:error-at-10}
\end{equation}
where $e_i=1$ indicates that the positive item $y_i$ is absent from the fixed
top-10 candidate set $B_i^{(10)}$. Error@10 concerns a single observed positive item and is narrower than overall ranking reliability; our monitoring conclusions are specific to this event. Let $u_i^{(m)}$ denote the scalar
uncertainty score produced by UQ method $m$, with larger values indicating
greater estimated uncertainty. Error-AUROC is our primary metric for measuring
how well $u_i^{(m)}$ separates Error@10=1 cases from records whose positive item
remains in the fixed top-10 ranking \citep{fawcett2006roc}. On the shared Warm-start records, we additionally report Error-AUPRC,
Risk@80, AURC, and E-AURC. Error-AUPRC complements AUROC when Error@10
cases are imbalanced. Risk@80 measures the Error@10 rate after discarding
the 20\% most uncertain records and retaining the remaining 80\%.
AURC summarizes this risk--coverage trade-off across all coverage levels,
while E-AURC measures the excess AURC above the oracle ordering
\citep{geifman2017selective}. Lower values are better for Risk@80, AURC,
and E-AURC.

\paragraph{Paired-Control Response Agreement.}
We further evaluate how UQ signals respond to controlled changes in user-history and item-content evidence.
The user-history control compares full and truncated histories and the item-content control compares original and ablated item text. Each signed NDCG@10 difference is the original-input score minus the mean score across the corresponding controlled variants. Negative differences indicate higher ranking quality under the
controlled inputs.
Let $N(\cdot)$ denote NDCG@10 \citep{jarvelin2002cumulated}.
For case $i$, the clipped NDCG@10 differences are
\begin{align}
    \Delta_{i,u}
    &=
    \left[
        N_{i,\mathrm{full}}
        -
        \overline{N}_{i,\mathrm{trunc}}
    \right]_{+},
    \label{eq:user-restoration-gain}\\
    \Delta_{i,v}
    &=
    \left[
        N_{i,\mathrm{orig}}
        -
        \overline{N}_{i,\mathrm{abl}}
    \right]_{+},
    \label{eq:item-restoration-gain}
\end{align}
where $[x]_{+}=\max(x,0)$, $u$ denotes user history, and $v$ denotes item
content. $N_{i,\mathrm{full}}$ and $N_{i,\mathrm{orig}}$ denote NDCG@10
under the full-history and original-item-text settings, while
$\overline{N}_{i,\mathrm{trunc}}$ and $\overline{N}_{i,\mathrm{abl}}$
denote averages over the corresponding truncated-history and item-text
ablation variants. Thus, $\Delta_{i,u}$ and $\Delta_{i,v}$ are the clipped
NDCG@10 differences for the user-history and item-content controls,
respectively. Cases with $\Delta_{i,u}=\Delta_{i,v}$ are excluded from the dominant-control evaluation. For the remaining cases, the operational
dominant-control label is
\begin{equation}
    d_i
    =
    \arg\max_{s\in\{u,v\}}
    \Delta_{i,s}.
    \label{eq:dominant-control}
\end{equation}
This label selects the source with the larger clipped NDCG@10 difference for case $i$. Among 4,000 Llama cases, history/item effects are negative in 49.5\%/38.8\%; clipping creates 1,798 additional ties and preserves the user/item labels of the remaining non-tied cases. The complementary signed analysis selects the larger unclipped effect (3,617 non-tied cases) and uses unclipped scalar responses (Appendix~\ref{sec:appendix-signed}). Let $r_{i,u}^{(m)}$ and $r_{i,v}^{(m)}$ denote the user-history and
item-content response scores of method $m$. Methods with
explicit control-specific responses use these scores directly. For
scalar-only methods, the two responses are derived from the increase in
uncertainty under the corresponding paired control, as detailed in
Appendix~\ref{sec:appendix-source-response}. For a non-tied response pair,
method $m$ predicts
\begin{equation}
    \hat{d}_i^{(m)}
    =
    \arg\max_{s\in\{u,v\}}
    r_{i,s}^{(m)}.
    \label{eq:predicted-control}
\end{equation}
The prediction therefore selects the control to which method $m$ gives the
larger response; equal response scores are treated as tie predictions. We report directional coverage, dominant-control accuracy, Macro-F1,
Source-AUROC, and Spearman rank correlation with the NDCG@10 differences. Directional coverage measures how often a method produces a non-tied
response. The remaining metrics measure how closely the method responses
agree with the relative NDCG@10 differences under the two constructed
controls.

\subsection{UQ Monitoring Signals}
\label{sec:method-uq-definitions}

Section~\ref{sec:method-evaluation} defines the evaluation targets and metrics
in terms of generic scalar uncertainty scores and control-specific response
scores. We next describe the eight UQ signals used to produce these quantities.
These signals span probability-based confidence, calibration, decomposition,
perturbation sensitivity, and sampling
\citep{gawlikowski2023survey,kweon2025uncertainty}. The resulting
ranking-level signals may reflect both aleatoric and epistemic uncertainty,
including ambiguity or incompleteness in user or item evidence and sensitivity
to the model, prompt, or sampling procedure. We evaluate their observed
monitoring behavior under a shared protocol without imposing an
aleatoric--epistemic decomposition. All signals monitor the same fixed base
ranking. Detailed score definitions, orientation conventions, and implementation
parameters are provided in Appendix~\ref{sec:appendix-uq-definitions}.

\paragraph{Probability and Calibration Signals.}
LabelProb-Max derives uncertainty from the maximum next-token probability over
the valid candidate labels \textit{A--T}
\citep{hendrycks2017baseline}. LabelProb-Entropy measures dispersion over the
same option-label distribution
\citep{shannon1948mathematical,settles2009active}, while LabelProb-Margin
measures ambiguity between the two most probable labels
\citep{scheffer2001active,settles2009active}. Temperature-scaled LabelProb
applies post-hoc scalar calibration to this distribution
\citep{guo2017calibration}. LabelProb-Top10Mass derives uncertainty from the
probability mass assigned to the fixed base top-10 set, thereby aligning the
probability-mass cutoff with the monitored Error@10 event. Top10Mass combines the option-label distribution from a separate
one-letter prompt with the issued listwise ranking, so it partly measures
cross-elicitation consistency between the two elicitation formats.
Appendix~\ref{sec:appendix-same-inference} compares it with teacher-forced
scores reconstructed under the original listwise ranking prompt.

\paragraph{Decomposition and Sampling Signals.}
Decomposed LLM UQ follows \citet{kweon2025uncertainty} and uses
candidate-order variants to separate the recommendation and prompt uncertainty components. Self-consistency draws five stochastic rankings for each record and measures
their top-10 disagreement, following prior work which uses output consistency as uncertainty evidence
\citep{wang2023selfconsistency,xiao2025consistency}.

\paragraph{Paired-Control Perturbation Signal.}
Perturbation Rank Instability measures how strongly the fixed top-10 set
changes under the constructed user-history and item-content controls,
following perturbation-based behavioral evaluation
\citep{ribeiro2016lime,ribeiro2020checklist}. In addition to its scalar
instability score, the method retains separate user-history and item-content
responses for the paired-control evaluation. Beyond the base ranking, the four LabelProb signals share one forward pass; temperature scaling reuses this pass and adds an offline calibration fit. Decomposed LLM UQ uses five forward passes, self-consistency five sampled generations, and Perturbation Rank Instability 7.3 controlled greedy generations on average. The Llama evaluation reuses the controlled rankings generated for the robustness conditions (Appendix~\ref{sec:appendix-cost}).

\subsection{Statistical Protocol}
\label{sec:method-statistics}

The overall score for each evaluation condition is computed by pooling all
eligible records in that condition. Warm-start confidence intervals use
1,000 bootstrap resamples stratified by dataset and paired across methods.
Sparse-history, Item-text robustness, and the tail-target subset use 200
paired bootstrap resamples stratified by dataset. For conditions containing
multiple variants of the same base instance, those variants are resampled
together. Implementation details, artifact alignment, parsing checks, and coverage
statistics are reported in Appendix~\ref{sec:appendix-repro}.

\section{Experiments and Analysis}
\label{sec:experiments}

We organize the fixed-ranking evaluation around eight observations.
Observation~1 examines backbone sensitivity, Observations~2--7 examine
ranking-failure detection, and Observation~8 examines paired-control
response agreement. Figure~\ref{fig:backbone-view} summarizes the main trends.
Additional diagnostics are reported in Appendix~\ref{sec:appendix-repro}.

\subsection{Experimental Settings}
\label{sec:exp-settings}

The full Llama-3.1-8B-Instruct evaluation suite contains 42,475 eligible ranking records: 4,000 Warm-start records, 20,000 Sparse-history records, 1,312 records in the tail-target subset, and 17,163 Item-text robustness records. The Qwen3-8B non-thinking evaluation covers only the same 4,000 Warm-start instances and tests whether Warm-start method ordering
reproduces on a second backbone. Warm-start is the reference condition. Sparse-history and Item-text robustness provide the user-history and item-content controls, respectively. The tail-target condition is formed by filtering Warm-start records by target popularity. Confidence intervals and pooled scores follow Section~\ref{sec:method-statistics}. Additional implementation, coverage, and evaluator details are reported in Appendix~\ref{sec:appendix-repro}.

\subsection{Warm-Start}
\label{sec:exp-warm-start}

\begin{figure*}[t]
\centering
\includegraphics[width=0.98\textwidth]
{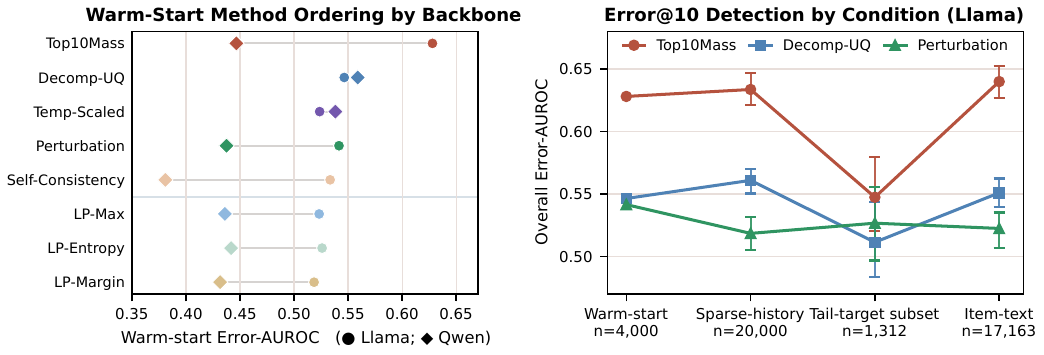}
\caption{\textbf{Backbone- and condition-specific ranking-failure detection.}
Warm-start Error-AUROC across backbones and aggregate Llama Error-AUROC across evaluation conditions.}
\label{fig:backbone-view}
\vspace{-2mm}
\end{figure*}

Warm-start serves as the reference condition, with the original user histories
and item descriptions left unchanged. To examine backbone sensitivity, we
evaluate the same 4,000 Warm-start instances with Llama-3.1-8B-Instruct and
Qwen3-8B in non-thinking mode. Table~\ref{tab:uq-warm-start} reports Error-AUROC
across datasets and backbones, while the left panel of
Figure~\ref{fig:backbone-view} summarizes the aggregate method ordering.

\begin{table*}[!t]
\centering
\footnotesize
\setlength{\tabcolsep}{3.5pt}
\renewcommand{\arraystretch}{1.08}
\setlength{\aboverulesep}{0.20ex}
\setlength{\belowrulesep}{0.30ex}
\begin{tabular}{lcccccccc}
\toprule
\rowcolor{panelshade}

& \shortstack{LabelProb\\Max}
& \shortstack{LabelProb\\Top10Mass}
& \shortstack{LabelProb\\Entropy}
& \shortstack{LabelProb\\Margin}
& \shortstack{Temperature\\scaling}
& \shortstack{Decomposed\\LLM UQ}
& \shortstack{Perturbation\\Rank Instability}
& \shortstack{Self-\\consistency} \\
\midrule
\multicolumn{9}{l}{\textit{Llama-3.1-8B-Instruct}} \\
\rowcolor{tablerow}
MIND-20
& 0.476 & \textbf{0.528} & 0.486 & 0.482
& 0.477 & 0.484 & \underline{0.517} & \textbf{0.528} \\

Books
& 0.607 & \textbf{0.673} & 0.601 & 0.583
& 0.608 & 0.620 & \underline{0.631} & 0.570 \\

\rowcolor{tablerow}
Movies
& 0.449 & \textbf{0.579} & 0.448 & 0.463
& 0.449 & 0.511 & 0.525 & \underline{0.529} \\

Yelp
& 0.551 & \textbf{0.686} & 0.552 & 0.540
& 0.551 & \underline{0.556} & 0.525 & 0.521 \\

\rowcolor{tablerow}
Overall
& 0.523 & \textbf{0.628} & 0.526 & 0.519
& 0.524 & \underline{0.546} & 0.542 & 0.533 \\
\midrule
\multicolumn{9}{l}{\textit{Qwen3-8B non-thinking}} \\
\rowcolor{tablerow}
MIND-20
& 0.475 & 0.486 & 0.476 & 0.472
& \underline{0.492} & \textbf{0.518} & 0.462 & 0.441 \\

Books
& 0.446 & 0.458 & 0.449 & 0.443
& \underline{0.558} & \textbf{0.576} & 0.509 & 0.493 \\

\rowcolor{tablerow}
Movies
& 0.439 & 0.426 & 0.440 & 0.437
& \underline{0.478} & \textbf{0.498} & 0.413 & 0.372 \\

Yelp
& 0.333 & 0.427 & 0.337 & 0.328
& \underline{0.487} & \textbf{0.567} & 0.400 & 0.363 \\

\rowcolor{tablerow}
Overall
& 0.436 & 0.447 & 0.442 & 0.432
& \underline{0.538} & \textbf{0.559} & 0.437 & 0.381 \\
\bottomrule
\end{tabular}
\caption{\textbf{Warm-start Error-AUROC across datasets and backbones.}
Results are computed on the same 4,000 Warm-start instances for both
backbones. Bold and underlined values indicate the best and second-best
methods in each row.}
\label{tab:uq-warm-start}
\end{table*}

\begin{table*}[!t]
\centering
\footnotesize
\setlength{\tabcolsep}{9pt}
\renewcommand{\arraystretch}{1.10}
\setlength{\aboverulesep}{0pt}
\setlength{\belowrulesep}{0pt}
\setlength{\extrarowheight}{1pt}
\begin{tabular}{lrrrr}
\toprule
\rowcolor{panelshade}
\rule[-0.7ex]{0pt}{3.2ex}
Method
& Error-AUPRC $\uparrow$
& Risk@80 $\downarrow$
& AURC $\downarrow$
& E-AURC $\downarrow$ \\
\midrule
\multicolumn{5}{l}{\textit{Llama-3.1-8B-Instruct}} \\
\rowcolor{tablerow}
LabelProb-Max
& 0.407 & 0.387 & 0.373 & 0.283 \\

LabelProb-Top10Mass
& \textbf{0.509} & \textbf{0.355}
& \textbf{0.298} & \textbf{0.208} \\

\rowcolor{tablerow}
LabelProb-Entropy
& 0.409 & 0.388 & 0.373 & 0.282 \\

LabelProb-Margin
& 0.401 & 0.392 & 0.371 & 0.281 \\

\rowcolor{tablerow}
Temperature-scaled LabelProb
& 0.408 & 0.385 & 0.373 & 0.283 \\

Decomposed LLM UQ
& 0.427 & 0.386 & \underline{0.355} & \underline{0.264} \\

\rowcolor{tablerow}
Perturbation Rank Instability
& \underline{0.435} & \underline{0.373} & 0.372 & 0.282 \\

Self-consistency
& 0.408 & 0.385 & 0.365 & 0.274 \\
\midrule
\multicolumn{5}{l}{\textit{Qwen3-8B non-thinking}} \\
\rowcolor{tablerow}
LabelProb-Max
& 0.321 & 0.391 & 0.386 & 0.308 \\

LabelProb-Top10Mass
& 0.319 & 0.409 & 0.372 & 0.295 \\

\rowcolor{tablerow}
LabelProb-Entropy
& 0.329 & 0.389 & 0.384 & 0.307 \\

LabelProb-Margin
& 0.318 & 0.392 & 0.389 & 0.312 \\

\rowcolor{tablerow}
Temperature-scaled LabelProb
& \underline{0.393} & \textbf{0.357}
& \underline{0.325} & \underline{0.247} \\

Decomposed LLM UQ
& \textbf{0.402} & \textbf{0.357}
& \textbf{0.307} & \textbf{0.230} \\

\rowcolor{tablerow}
Perturbation Rank Instability
& 0.352 & \underline{0.369} & 0.423 & 0.346 \\

Self-consistency
& 0.329 & 0.383 & 0.448 & 0.370 \\
\bottomrule
\end{tabular}
\caption{\textbf{Warm-start results on complementary reliability metrics.}
Error-AUPRC, Risk@80, AURC, and E-AURC are reported on the same 4,000
instances for both backbones. Bold and underlined values indicate the best
and second-best methods within each backbone block.}
\label{tab:uq-warm-multimetric}
\vspace{-4mm}
\end{table*}

\noindent\textbf{\textit{Observation 1: UQ method ordering is
backbone-dependent.}}
Top10Mass ranks first on Llama but falls to third on Qwen, while Decomposed
LLM UQ moves from second to first and temperature scaling moves from sixth to
second. Moreover, the Top10Mass minus Decomposed Error-AUROC margin reverses sign, from $+0.082$ on Llama to $-0.112$ on Qwen.
Because the evaluated instances and protocol are held fixed, this reversal
shows that the relative effectiveness of UQ monitors can depend strongly on
the backbone. Several Qwen monitors also yield Error-AUROC values below $0.5$, indicating
that their predefined uncertainty scores are inversely associated with
Error@10 in this setting. Appendix~\ref{sec:appendix-sanity} provides
additional score-orientation and error-versus-correct diagnostics. We further evaluate the same Warm-start records with complementary reliability
metrics. Table~\ref{tab:uq-warm-multimetric} reports Error-AUPRC, Risk@80,
AURC, and E-AURC for both backbones. Appendix Table~\ref{tab:appendix-topk-alignment} provides an additional check of whether alignment between the monitored cutoff and the probability-mass cutoff is associated with Error@K detection.

\noindent\textbf{\textit{Observation 2: Within the TopKMass family,
matching the cutoff to Error@K improves detection.}}
We compare TopKMass variants with $K=5,10,15$ to examine cutoff alignment. Top5Mass, Top10Mass, and Top15Mass respectively perform best for Error@5, Error@10, and Error@15. Their advantages over the strongest mismatched cutoff
are $0.022$, $0.012$, and $0.036$ AUROC (Appendix Table~\ref{tab:appendix-topk-alignment}). On Llama Warm-start, using the issued top-10 set improves Error-AUROC by 0.103 over the probe's own top-10 set under the same one-letter distribution.
The reconstructed first-slot top-10-mass score under the original ranking prompt is 0.022 lower than the cross-elicitation Top10Mass score (Appendix~\ref{sec:appendix-same-inference}). The issued-set comparison supports cutoff-aligned monitoring.

\subsection{Sparse-History}
\label{sec:exp-sparse-history}

Sparse-history evaluates UQ monitoring after truncating the available user
history while keeping the candidate set and ranking protocol unchanged.
Table~\ref{tab:uq-sparse-history} reports the dataset-level and pooled
Error-AUROC and the right panel of Figure~\ref{fig:backbone-view}
summarizes the aggregate Llama result. Instance-clustered confidence intervals
are reported in Appendix Table~\ref{tab:appendix-cluster-bootstrap}.

\begin{table*}[!t]
\centering
\footnotesize
\setlength{\tabcolsep}{3.5pt}
\renewcommand{\arraystretch}{0.96}
\setlength{\aboverulesep}{0.20ex}
\setlength{\belowrulesep}{0.30ex}
\begin{tabular}{lcccccccc}
\toprule
\rowcolor{panelshade}
Dataset
& \shortstack{LabelProb\\Max}
& \shortstack{LabelProb\\Top10Mass}
& \shortstack{LabelProb\\Entropy}
& \shortstack{LabelProb\\Margin}
& \shortstack{Temperature\\scaling}
& \shortstack{Decomposed\\LLM UQ}
& \shortstack{Perturbation\\Rank Instability}
& \shortstack{Self-\\consistency} \\
\midrule
\rowcolor{tablerow}
MIND-20
& 0.488 & \underline{0.515} & 0.485 & 0.493
& 0.487 & 0.484 & \textbf{0.516} & 0.509 \\

Books
& 0.610 & \textbf{0.696} & 0.608 & 0.581
& 0.611 & \underline{0.630} & 0.558 & 0.584 \\

\rowcolor{tablerow}
Movies
& 0.455 & \textbf{0.575} & 0.454 & 0.476
& 0.455 & 0.515 & 0.519 & \underline{0.533} \\

Yelp
& \underline{0.596} & \textbf{0.694} & 0.594 & 0.571
& \underline{0.596} & 0.586 & 0.506 & 0.549 \\

\rowcolor{tablerow}
Overall
& 0.536 & \textbf{0.634} & 0.536 & 0.532
& 0.536 & \underline{0.561} & 0.519 & 0.536 \\
\bottomrule
\end{tabular}
\caption{\textbf{Sparse-history UQ results.}
Error-AUROC for Error@10 detection under user-history truncation on Llama-3.1-8B-Instruct.}
\label{tab:uq-sparse-history}
\vspace{-4mm}
\end{table*}

\noindent\textbf{\textit{Observation 3: Top10Mass remains the strongest
pooled monitor under Sparse-history.}}
Top10Mass reaches a pooled Error-AUROC of $0.634$, exceeding Decomposed LLM UQ by $0.073$, with non-overlapping 95\% instance-clustered bootstrap confidence intervals. It also ranks first on
Books, Movies, and Yelp. At the same time, Top10Mass trails Perturbation Rank Instability by only
$0.001$ on MIND-20. Because Top10Mass leads on three of four datasets and is nearly tied for first on the
fourth, its pooled advantage is not driven by a single domain. This pattern
suggests that the ranking-aligned Top10Mass signal remains informative for
Error@10 detection even when the available user-history evidence is reduced.

\subsection{Tail-Target Subset}
\label{sec:exp-long-tail}

The tail-target subset evaluates the Warm-start records whose positive item
falls in the dataset-specific popularity-tail bucket. It contains 1,312 records with the retained counts varying across datasets.
Table~\ref{tab:uq-tail-target} reports dataset-level and pooled Error-AUROC and the right panel of Figure~\ref{fig:backbone-view} summarizes the
aggregate result. Confidence intervals are reported in
Appendix Table~\ref{tab:appendix-cluster-bootstrap}.

\begin{table*}[!t]
\centering
\footnotesize
\setlength{\tabcolsep}{3.0pt}
\renewcommand{\arraystretch}{0.96}
\setlength{\aboverulesep}{0.20ex}
\setlength{\belowrulesep}{0.30ex}
\begin{tabular}{lrcccccccc}
\toprule
\rowcolor{panelshade}
Dataset & $n$
& \shortstack{LabelProb\\Max}
& \shortstack{LabelProb\\Top10Mass}
& \shortstack{LabelProb\\Entropy}
& \shortstack{LabelProb\\Margin}
& \shortstack{Temperature\\scaling}
& \shortstack{Decomposed\\LLM UQ}
& \shortstack{Perturbation\\Rank Instability}
& \shortstack{Self-\\consistency} \\
\midrule
\rowcolor{tablerow}
MIND-20 & 870
& 0.463 & 0.503 & 0.477 & 0.471
& 0.464 & 0.483 & \textbf{0.525} & \underline{0.505} \\

Books & 221
& 0.575 & \textbf{0.643} & 0.588 & 0.560
& 0.578 & 0.587 & \underline{0.598} & 0.492 \\

\rowcolor{tablerow}
Movies & 129
& 0.522 & \underline{0.529} & 0.513 & \textbf{0.540}
& 0.522 & 0.482 & 0.411 & 0.485 \\

Yelp & 92
& 0.651 & \textbf{0.687} & 0.608 & \underline{0.675}
& 0.647 & 0.630 & 0.558 & 0.585 \\

\rowcolor{tablerow}
Overall & 1,312
& 0.503 & \textbf{0.547} & 0.510 & 0.511
& 0.504 & 0.511 & \underline{0.527} & 0.506 \\
\bottomrule
\end{tabular}
\caption{\textbf{Tail-target subset UQ results on Llama-3.1-8B-Instruct.}
Error-AUROC for Error@10 detection on the 1,312 Warm-start records whose
target item belongs to the dataset-specific popularity-tail bucket.}
\label{tab:uq-tail-target}
\vspace{-2mm}
\end{table*}

\noindent\textbf{\textit{Observation 4: Popularity-tail targets weaken the
Warm-start advantage of Top10Mass.}}
Top10Mass reaches a pooled Error-AUROC of $0.547$, which is $0.081$ lower
than its Warm-start score. Its pooled lead over Perturbation Rank Instability is only $0.020$, and their instance-clustered confidence intervals overlap. These results show that the strong Warm-start advantage of Top10Mass
does not extend uniformly to popularity-tail targets. 

\noindent\textbf{\textit{Observation 5: Tail-target UQ performance varies across datasets.}}
The dataset-level winner varies across domains. Perturbation Rank Instability
leads on MIND-20, LabelProb-Margin leads on Movies, while Top10Mass remains
best on Books and Yelp. The pooled estimate should therefore be interpreted
alongside the dataset-level results, as MIND-20 contributes $870$ of the
1,312 retained records.

\subsection{Item-Text Robustness}
\label{sec:exp-item-text}

Item-text robustness evaluates UQ monitoring under controlled candidate
item-text field ablations while keeping the remaining ranking protocol
unchanged. Table~\ref{tab:uq-item-text} reports dataset-level and pooled Error-AUROC. The right panel of Figure~\ref{fig:backbone-view} summarizes the aggregate
Llama result. The corresponding item-content NDCG@10 difference is evaluated
separately in the paired-control analysis.

\begin{table*}[!t]
\centering
\footnotesize
\setlength{\tabcolsep}{3.5pt}
\renewcommand{\arraystretch}{0.96}
\setlength{\aboverulesep}{0.20ex}
\setlength{\belowrulesep}{0.30ex}
\begin{tabular}{lcccccccc}
\toprule
\rowcolor{panelshade}
Dataset
& \shortstack{LabelProb\\Max}
& \shortstack{LabelProb\\Top10Mass}
& \shortstack{LabelProb\\Entropy}
& \shortstack{LabelProb\\Margin}
& \shortstack{Temperature\\scaling}
& \shortstack{Decomposed\\LLM UQ}
& \shortstack{Perturbation\\Rank Instability}
& \shortstack{Self-\\consistency} \\
\midrule
\rowcolor{tablerow}
MIND-20
& 0.488 & \textbf{0.521} & 0.492 & 0.490
& 0.488 & 0.484 & \underline{0.519} & 0.502 \\

Books
& 0.577 & \textbf{0.663} & 0.567 & 0.566
& 0.578 & \underline{0.597} & 0.586 & 0.553 \\

\rowcolor{tablerow}
Movies
& 0.463 & \textbf{0.618} & 0.452 & 0.476
& 0.462 & 0.509 & 0.507 & \underline{0.532} \\

Yelp
& \underline{0.564} & \textbf{0.677} & 0.554 & 0.548
& 0.563 & 0.559 & 0.506 & 0.519 \\

\rowcolor{tablerow}
Overall
& 0.537 & \textbf{0.640} & 0.534 & 0.530
& 0.537 & \underline{0.551} & 0.523 & 0.523 \\
\bottomrule
\end{tabular}
\caption{\textbf{Item-text robustness on Llama-3.1-8B-Instruct.}
Error-AUROC for Error@10 detection under controlled candidate item-text
field ablations.}
\label{tab:uq-item-text}

\end{table*}

\noindent\textbf{\textit{Observation 6: Top10Mass remains the strongest
pooled monitor under item-text ablation.}}
Top10Mass reaches a pooled Error-AUROC of $0.640$, exceeding Decomposed LLM UQ by $0.089$, with non-overlapping 95\% instance-clustered bootstrap
confidence intervals. 

\noindent\textbf{\textit{Observation 7: The item-text robustness of Top10Mass is consistent across datasets.}}
Top10Mass ranks first on all four datasets, leading the strongest alternative
by $0.066$ on Books, $0.086$ on Movies, and $0.113$ on Yelp. MIND-20 is the only near-tie, where its lead over Perturbation Rank Instability is $0.002$. This pattern shows that the pooled advantage is supported by multiple domains
and is not driven by a single high-performing dataset. Although the magnitude of the advantage varies across domains, its direction remains consistent under the controlled item-text ablations.

\subsection{Paired-Control Response}
\label{sec:exp-source-response}

The paired-control evaluation compares full and truncated histories
for the user-history control, and original and ablated item text for the
item-content control. We evaluate 1,819 cases with unequal clipped
NDCG@10 differences. Table~\ref{tab:source-response} reports all five agreement metrics. Table~\ref{tab:source-response} additionally reports signed-effect
agreement and simple non-UQ ranking-distance baselines. Additional visualization and sensitivity analyses are provided in
Appendix~\ref{sec:appendix-robustness-checks}. As defined in Section~\ref{sec:method-evaluation}, the analysis is
scoped to these two constructed controls.

\begin{table*}[t]
\centering
\footnotesize
\setlength{\tabcolsep}{3.6pt}
\renewcommand{\arraystretch}{1.08}
\setlength{\aboverulesep}{0pt}
\setlength{\belowrulesep}{0pt}
\setlength{\extrarowheight}{1pt}

\begin{tabular}{llrrrrrrr}
\toprule

\rowcolor{panelshade}
& & & & & & &
\multicolumn{2}{c}{Signed effects} \\
\cmidrule(lr){8-9}

\rowcolor{panelshade}
\rule[-0.7ex]{0pt}{3.2ex}
Method & Signal & DirCov & D-Acc & Macro-F1 & Source-AUROC & $\rho$
& D-Acc & Src-AUROC \\

\midrule

\rowcolor{tablerow}
Decomposed LLM UQ
& scalar resp.
& 0.487 & 0.267 & 0.328 & 0.521 & 0.071
& \textbf{0.579} & \textbf{0.548} \\

LabelProb-Max
& scalar resp.
& 0.482 & 0.262 & 0.331 & 0.520 & 0.054
& 0.570 & 0.535 \\

\rowcolor{tablerow}
LabelProb-Top10Mass
& scalar resp.
& 0.656 & 0.374 & 0.447 & 0.531 & 0.092
& 0.571 & 0.531 \\

LabelProb-Entropy
& scalar resp.
& 0.478 & 0.250 & 0.301 & 0.517 & 0.060
& 0.571 & 0.539 \\

\rowcolor{tablerow}
LabelProb-Margin
& scalar resp.
& 0.500 & 0.268 & 0.347 & 0.518 & 0.040
& 0.551 & 0.527 \\

Temperature-scaled LabelProb
& scalar resp.
& 0.479 & 0.263 & 0.332 & 0.521 & 0.056
& 0.572 & 0.535 \\

\rowcolor{tablerow}
Self-consistency
& scalar resp.
& 0.619 & 0.316 & 0.377 & 0.503 & -0.006
& 0.529 & 0.506 \\

Perturbation Rank Instability
& source scores
& \textbf{0.987} & \textbf{0.709} & \textbf{0.712}
& \textbf{0.675} & \textbf{0.456}
& 0.482 & 0.496 \\

\midrule

\rowcolor{tablerow}
Top-5 Jaccard distance
& ranking dist.
& 0.977 & 0.700 & 0.706
& 0.671 & 0.440
& 0.475 & 0.498 \\

Kendall distance
& ranking dist.
& 0.997 & 0.725 & 0.725
& 0.692 & 0.498
& 0.482 & 0.504 \\

\rowcolor{tablerow}
RBO distance ($p=0.9$)
& ranking dist.
& 1.000 & 0.746 & 0.744
& 0.692 & 0.494
& 0.497 & 0.505 \\

Majority class
& reference
& 1.000 & 0.502 & 0.334
& 0.500 & --
& 0.575 & 0.500 \\

\bottomrule
\end{tabular}

\caption{\textbf{Paired-control response on Llama-3.1-8B-Instruct.}
Clipped metrics use 1,819 non-tied cases; signed metrics use 3,617.
Bold indicates the best UQ signal in each column.}
\label{tab:source-response}
\vspace{-2mm}
\end{table*}

\noindent\textbf{\textit{Observation 8: Paired-control agreement reflects source-specific ranking sensitivity.}}
Perturbation Rank Instability leads the UQ signals on all five
clipped response metrics. Compared with Top10Mass, the strongest
scalar response, it improves directional coverage by $0.331$ and Source-AUROC by $0.144$. When restricting the evaluation to
cases with an absolute gap between the two clipped NDCG@10 differences above $0.10$,
dominant-control accuracy increases by $0.066$ and Source-AUROC by $0.064$,
while directional coverage increases by only $0.002$
(Table~\ref{tab:appendix-source-gap}). Perturbation Rank Instability uses the same user-history and
item-content controls that define the paired-control labels.
This alignment can favor the method on this task.
Scalar UQ methods derive both responses from changes in a single
uncertainty score. The same qualitative conclusion holds when restricting the analysis to cases
with strictly positive NDCG@10 differences for both controls. Kendall and RBO each improve clipped Source-AUROC by 0.017 over
Perturbation Rank Instability, supporting source-specific ranking
sensitivity. For signed effects, the ranking-distance signals remain
near chance in Source-AUROC. The best scalar signal exceeds the
majority reference by 0.004 in D-Acc.

\subsection{Case Study and Diagnostic Checks}
\label{sec:exp-diagnostics}

\begin{figure}[t]
\centering
\includegraphics[width=0.94\columnwidth]
{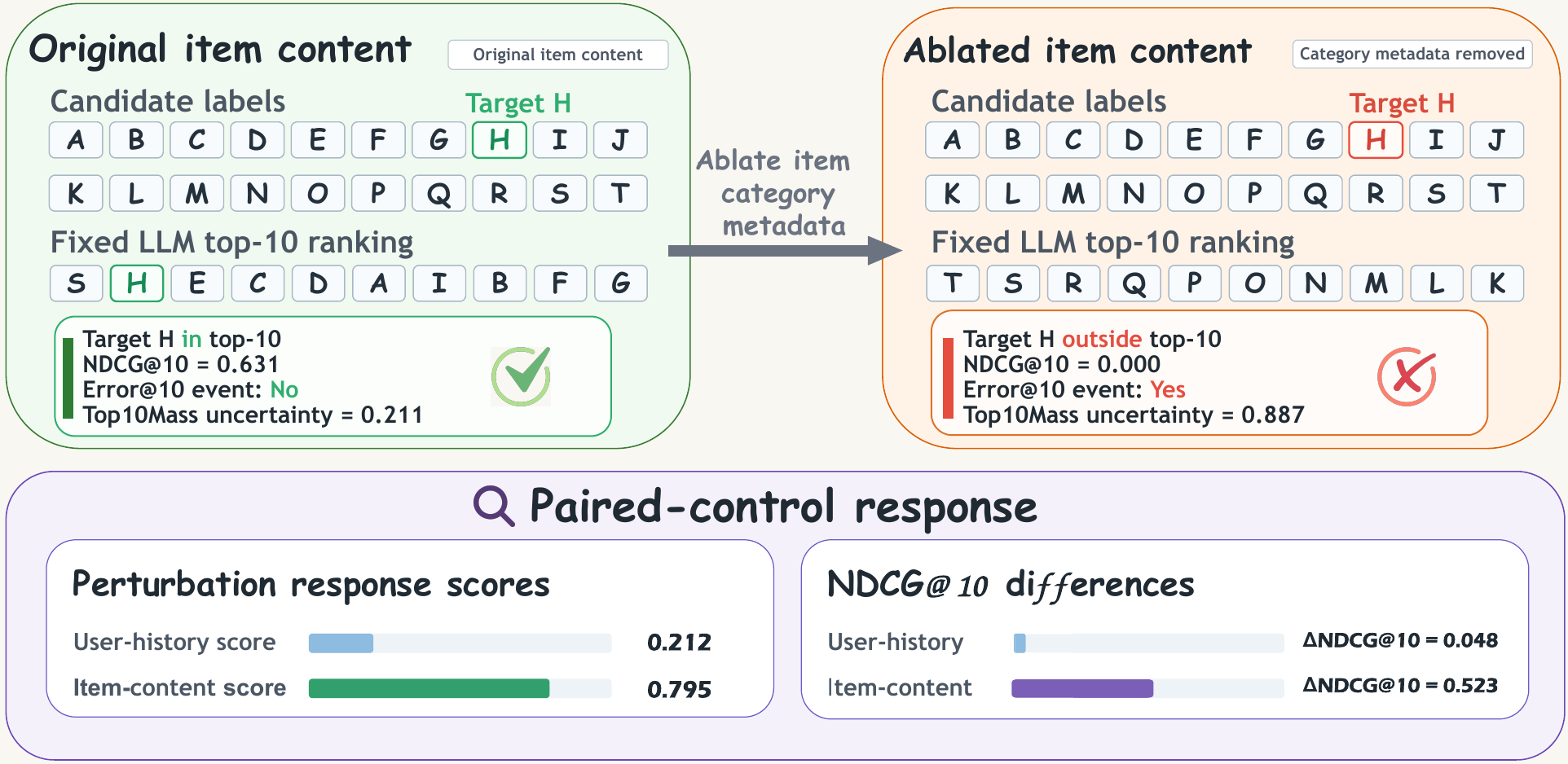}
\caption{\textbf{Illustrative item-content case.}
Removing category metadata moves the target outside the fixed top-10; both
the perturbation response and NDCG@10 difference are larger for item content.}
\label{fig:case-study}
\vspace{-4mm}
\end{figure}

\noindent\textbf{$\blacktriangleright$ Illustrative Case Study.}
Figure~\ref{fig:case-study} connects the scalar error monitoring with the paired-control response on one Amazon Books example. Removing the category metadata moves the target outside the fixed top-10 ranking, decreases NDCG@10 by 0.631, and increases Top10Mass uncertainty by 0.676. The item-content perturbation response exceeds the user-history response by 0.583.
The item-content NDCG@10 difference exceeds the user-history
difference by 0.475.
The case therefore provides a concrete instance of the aggregate paired-control
pattern in Observation~8, where the stronger perturbation response agrees with
the source with the larger clipped NDCG@10 difference.

\smallskip
\noindent\textbf{$\blacktriangleright$ Sanity and Component Checks.}
Tables~\ref{tab:appendix-sanity} and~\ref{tab:appendix-derived-checks} report the sanity and component checks over all 42,475 eligible Llama records. Random scoring and the parse-failure indicator remain near chance, while the oracle reaches 1.000. Top10Mass exceeds LabelProb-BaseTop1 by 0.050 in Error-AUROC. The decomposition-prompt and self-consistency top-1 components remain near chance.

\section{Conclusion}
\label{sec:conclusion}

We present a fixed-ranking benchmark for uncertainty quantification in
LLM-based recommendation, where UQ methods monitor the same issued rankings
without changing the candidate order. Across four datasets, the benchmark
evaluates ranking-failure detection, the stability of UQ method performance
across backbones and evaluation conditions, and paired-control response
agreement for user-history and item-content evidence. Our results show that no single UQ ordering is preserved across all settings. Top10Mass achieves the strongest pooled Error@10 detection performance
under several Llama conditions, reflecting effective cutoff-aligned monitoring, while its advantage becomes smaller on the tail-target subset. The same Warm-start evaluation on Qwen produces a different method ordering.
For paired controls, Perturbation Rank Instability provides stronger agreement
with the relative clipped NDCG@10 differences than the scalar UQ signals; simple listwise distances also agree strongly with these clipped differences, supporting source-specific diagnosis. These results
show that UQ quality is task-dependent. A scalar monitor that is effective for ranking-failure detection may not be the most informative for controlled evidence-response agreement. The practical monitoring claim is limited to Error@10 for a single observed positive item. A UQ monitor validated on the target backbone and recommendation
condition can route high-uncertainty requests to a fallback recommender.
Risk--coverage curves on held-out data can guide threshold selection.
Risk@80 and AURC describe ranking errors among retained LLM outputs.
An online A/B test of the complete routing policy would measure
click-through and conversion rates. By holding the issued ranking fixed and aligning
all methods to the same records, our benchmark makes these differences directly
comparable without conflating monitoring quality with reranking behavior.
This separation provides a common basis for evaluating UQ methods across recommendation conditions.

\section{Acknowledgment}

This work was partially supported by NSF 2616633, NSF 2303037, NSF 2312501, NSF 2531008, NSF 2553619, NSF 2617441, NSF 2609582, DARPA HR001126CE054, NIH OT2OD038003, NIH U54OD036472, NIH U54DK097771, Optum, NEC, SRC JUMP 2.0 Center, CDSC Industry Partners, Amazon Research Awards, Snapchat, and Google Gifts. The computing resources for this work were supported in part by the University of North Carolina at Chapel Hill through its AI Acceleration Program. Hang Zhou is partially supported by the SDSS Computational Data Science \& AI Cloud Pilot Program at the University of North Carolina at Chapel Hill.

\bibliography{custom}
\bibliographystyle{tmlr}

\appendix

\section{Reproducibility and Artifact Details}
\label{sec:appendix-repro}

This appendix provides the implementation details needed to reproduce the
fixed-ranking benchmark defined in Section~\ref{sec:method}. We describe the
artifact contract, backbone and sampling settings, exact UQ score definitions,
paired-control response construction, coverage checks, and supplementary
diagnostics.

\subsection{Artifact and Evaluator Contract}
\label{sec:appendix-artifacts}
\label{sec:appendix-evaluator-contract}

The benchmark stores the fixed base recommendation output separately from the
UQ outputs. The base prediction table contains the parsed 20-candidate ranking,
record-level recommendation metrics, parse status, and the binary label
\texttt{is\_error\_at\_10}. Each UQ artifact is aligned to the corresponding
base record through \texttt{record\_id}. The Error@10 label is used only by the
evaluator and is not used to construct a UQ score.

Table~\ref{tab:appendix-schema} summarizes the shared UQ artifact schema.

\begin{table}[t]
\centering
\setlength{\tabcolsep}{6pt}

\begin{tabular}{ll}
\toprule
Field & Description \\
\midrule
\texttt{schema\_version} & Artifact schema version. \\
\texttt{record\_id} & Stable join key for base and UQ rows. \\
\texttt{method} & UQ method identifier. \\
\texttt{method\_family} & Probability, calibration, decomposition, sensitivity, or sampling. \\
\texttt{uncertainty\_score} & Scalar score; larger means more uncertain. \\
\texttt{confidence\_score} & Optional confidence counterpart. \\
\texttt{source\_scores} & Optional paired-control user/item response scores. \\
\texttt{components} & Optional method-specific diagnostic values. \\
\texttt{metadata} & Optional run settings and parser status. \\
\bottomrule
\end{tabular}%

\caption{\textbf{UQ artifact schema.}
Required and optional fields for record-aligned UQ outputs.}
\label{tab:appendix-schema}
\end{table}

The scalar evaluator joins each UQ artifact to the fixed base table and
reports Error-AUROC for the primary Error@10 analysis. It also produces
Error-AUPRC, Risk@80, AURC, and E-AURC for the complementary reliability
analysis. Results can be summarized by method, evaluation condition, and
dataset. The paired-control evaluator computes directional coverage,
dominant-control accuracy, Macro-F1, Source-AUROC, and Spearman rank
correlation using the operational labels defined in
Section~\ref{sec:method-evaluation}. The diagnostics in
Appendices~\ref{sec:appendix-robustness-checks}--\ref{sec:appendix-derived-checks}
operate on existing artifacts and require no additional LLM inference.

\subsection{Backbone and Sampling Configuration}
\label{sec:appendix-backbone-config}
\label{sec:appendix-sampling-status}

Llama-3.1-8B-Instruct provides the full evaluation suite. Qwen3-8B is evaluated
in non-thinking mode on the shared 4,000 Warm-start records. Qwen generation
for this replication is deterministic. Model identifiers and available revision information, numerical
precision, seeds, and generation arguments are stored in the run metadata.
Perturbation Rank Instability may use auxiliary rankings as monitor inputs;
these auxiliary rankings do not constitute additional reported evaluation
conditions. Self-consistency draws five stochastic rankings per record with
temperature $0.9$ and top-$p$ $0.95$. The Llama sampled-ranking outputs have a
mean parse success rate of $0.995$. Because self-consistency requires repeated
generations, it has higher inference cost than the single-pass LabelProb
signals.

\subsection{Detailed UQ Signal Definitions}
\label{sec:appendix-uq-definitions}

Let record $i$ contain $K=20$ candidates with valid option labels
\begin{equation}
    \mathcal{A}_i=\{A,\ldots,T\}.
\end{equation}
All scalar UQ scores are oriented so that larger values indicate greater
uncertainty.

\paragraph{Option-Label Distribution.}
For probability-based methods, we use a one-letter choice prompt and read the
next-token logits of the valid option labels. Let $\mathcal{T}(a)$ denote the
supported token forms for label $a$, including bare and space-prefixed forms.
We define
\begin{equation}
    z_i(a)
    =
    \max_{t\in\mathcal{T}(a)}
    \operatorname{logit}_i(t),
\end{equation}
where $z_i(a)$ is the logit assigned to option label $a$ after selecting the
largest supported token form. We then normalize over the valid labels:
\begin{equation}
    p_i(a)
    =
    \frac{\exp z_i(a)}
    {\sum_{b\in\mathcal{A}_i}\exp z_i(b)}.
\end{equation}
Thus, $p_i(a)$ is the option-label probability used by the LabelProb family.

\paragraph{Probability Scores.}
LabelProb-Max uses
\begin{equation}
    u_i^{\mathrm{max}}
    =
    1-\max_{a\in\mathcal{A}_i}p_i(a).
\end{equation}
The score is high when no single valid option label receives high probability.

Let $\mathcal{A}_i^{(10)}$ denote the option labels corresponding to the fixed
candidate set $B_i^{(10)}$. LabelProb-Top10Mass uses
\begin{equation}
    u_i^{\mathrm{top10}}
    =
    1-
    \sum_{a\in\mathcal{A}_i^{(10)}}
    p_i(a).
\end{equation}
This score measures the probability mass outside the fixed base top-10 set. It
uses no relevance label and does not modify the base ranking.

LabelProb-Entropy uses normalized entropy,
\begin{equation}
    u_i^{\mathrm{ent}}
    =
    -\frac{1}{\log K}
    \sum_{a\in\mathcal{A}_i}
    p_i(a)\log p_i(a),
\end{equation}
which increases as probability mass becomes more dispersed across the valid
labels. LabelProb-Margin uses the inverted top-two margin,
\begin{equation}
    u_i^{\mathrm{margin}}
    =
    1-
    \left(
        p_{i,(1)}-p_{i,(2)}
    \right),
\end{equation}
where $p_{i,(1)}$ and $p_{i,(2)}$ are the largest and second-largest
option-label probabilities, respectively.

\paragraph{Temperature-Scaled LabelProb.}
Temperature scaling applies a scalar temperature $T>0$ to the option-label
distribution:
\begin{equation}
    p_i^{(T)}(a)
    =
    \frac{p_i(a)^{1/T}}
    {\sum_{b\in\mathcal{A}_i}p_i(b)^{1/T}}.
\end{equation}
We fit $T$ by cross-validation using the positive option label $a_i^{+}$:
\begin{equation}
    \mathcal{L}(T)
    =
    -\sum_i
    \log p_i^{(T)}(a_i^{+}).
\end{equation}
The calibrated uncertainty score is
\begin{equation}
    u_i^{\mathrm{temp}}
    =
    1-
    \max_{a\in\mathcal{A}_i}
    p_i^{(T)}(a).
\end{equation}
Five folds are split by base-instance identifier to avoid leakage across
related variants. The fitted fold temperatures are $1.199$--$1.245$ for Llama
and $8.563$--$8.717$ for Qwen; fold-specific values are stored in the run
metadata. Table~\ref{tab:appendix-shared-temperature} reports the four LabelProb scores
under the same three fixed temperatures on both backbones.
The Top10Mass--Decomposed LLM UQ reversal persists across the tested temperatures.
At $T=8.62$, Qwen's LabelProb-Max and LabelProb-Entropy gain 0.102 and 0.157
in Error-AUROC relative to $T=1$.
Llama's scores change little across these settings.

\paragraph{Decomposed LLM UQ.}
Following decomposition-based LLM recommendation UQ, we create $M=5$
candidate-order variants and align their option-label distributions back to the
original candidates. Let $p_{im}(c)$ denote the aligned distribution over
original candidate $c$ under prompt variant $m$, and define
\begin{equation}
    \bar{p}_i(c)
    =
    \frac{1}{M}
    \sum_{m=1}^{M}
    p_{im}(c).
\end{equation}
Using normalized entropy $H(\cdot)$, we compute
\begin{align}
    U_i^{\mathrm{total}}
    &=
    H(\bar{p}_i),\\
    U_i^{\mathrm{rec}}
    &=
    \frac{1}{M}
    \sum_{m=1}^{M}
    H(p_{im}),\\
    U_i^{\mathrm{prompt}}
    &=
    U_i^{\mathrm{total}}
    -
    U_i^{\mathrm{rec}}.
\end{align}
The main benchmark score is
\begin{equation}
    u_i^{\mathrm{decomp}}
    =
    U_i^{\mathrm{total}}.
\end{equation}
The recommendation and prompt components are retained for the component
diagnostics in Table~\ref{tab:appendix-derived-checks}.

\paragraph{Perturbation Rank Instability.}
For control $q$, let $R_{i,q}^{\mathrm{rest}}$ denote the reference ranking and
$R_{i,q}^{\mathrm{pert}}$ the corresponding perturbed ranking. Let $T_k(R)$
denote the top-$k$ candidate set of ranking $R$. We define the top-$k$ Jaccard
instability as
\begin{equation}
    d_{i,q}
    =
    1-
    \frac{
        \left|
            T_k(R_{i,q}^{\mathrm{rest}})
            \cap
            T_k(R_{i,q}^{\mathrm{pert}})
        \right|
    }{
        \left|
            T_k(R_{i,q}^{\mathrm{rest}})
            \cup
            T_k(R_{i,q}^{\mathrm{pert}})
        \right|
    }.
\end{equation}
A larger $d_{i,q}$ indicates a larger change in the top-$k$ recommendation
set under control $q$. We use $k=10$ and define the scalar perturbation score
as the mean available control-specific instability,
\begin{equation}
    u_i^{\mathrm{pert}}
    =
    \frac{1}{|\mathcal{Q}_i|}
    \sum_{q\in\mathcal{Q}_i}
    d_{i,q},
\end{equation}
where $\mathcal{Q}_i$ is the set of available paired controls for record $i$.
The artifact field \texttt{source\_scores} stores the separate user-history
and item-content responses used by the paired-control evaluation.

\paragraph{Self-Consistency Ranking Disagreement.}
The base ranking remains fixed while the method draws $S$ additional
stochastic rankings $\{R_{is}\}_{s=1}^{S}$. Self-consistency uncertainty is
the mean pairwise top-$k$ Jaccard disagreement among the sampled rankings:
\begin{equation}
    u_i^{\mathrm{sc}}
    =
    \binom{S}{2}^{-1}
    \sum_{s<t}
    \left(
        1-
        \frac{
            |T_k(R_{is})\cap T_k(R_{it})|
        }{
            |T_k(R_{is})\cup T_k(R_{it})|
        }
    \right).
    \label{eq:sc}
\end{equation}
We use $k=10$ and $S=5$. Larger values indicate less agreement among the
sampled top-10 sets. Pairwise sample disagreement measures dispersion of the model's
ranking distribution. Base-anchored disagreement measures how atypical
the issued ranking is within that distribution.
Appendix~\ref{sec:appendix-sc-base} evaluates the mean top-10 Jaccard
distance from the base ranking to each of the five samples and the
pairwise score over those samples plus the base ranking.

\begin{table}[t]
\centering
\small
\setlength{\tabcolsep}{5pt}
\begin{tabular}{lcccccc}
\toprule
 & \multicolumn{3}{c}{Llama-3.1-8B-Instruct} & \multicolumn{3}{c}{Qwen3-8B non-thinking} \\
\cmidrule(lr){2-4}\cmidrule(lr){5-7}
Score & $T{=}1$ & $T{=}1.23$ & $T{=}8.62$ & $T{=}1$ & $T{=}1.23$ & $T{=}8.62$ \\
\midrule
LabelProb-Max & 0.523 & 0.524 & 0.525 & 0.436 & 0.441 & 0.538 \\
LabelProb-Entropy & 0.526 & 0.526 & 0.526 & 0.442 & 0.451 & 0.599 \\
LabelProb-Margin & 0.519 & 0.519 & 0.519 & 0.432 & 0.434 & 0.458 \\
LabelProb-Top10Mass & 0.628 & 0.627 & 0.623 & 0.447 & 0.448 & 0.444 \\
\midrule
Decomposed LLM UQ & \multicolumn{3}{c}{0.546} & \multicolumn{3}{c}{0.559} \\
\bottomrule
\end{tabular}
\caption{\textbf{Warm-start Error-AUROC under shared fixed temperatures.}
The same $T$ is applied to both backbones.
$T=1.23$ and $T=8.62$ are the mean fitted temperatures for Llama and Qwen,
respectively. Decomposed LLM UQ is reported as an unchanged reference.}
\label{tab:appendix-shared-temperature}
\end{table}

\subsection{Paired-Control Response Construction}
\label{sec:appendix-source-response}

For scalar-only method $m$, the paired-control response scores are derived from
the increase in uncertainty under each control:
\begin{align}
    r_{i,u}^{(m)}
    &=
    \left[
        \bar{u}_{i,\mathrm{trunc}}^{(m)}
        -
        u_{i,\mathrm{full}}^{(m)}
    \right]_{+},\\
    r_{i,v}^{(m)}
    &=
    \left[
        \bar{u}_{i,\mathrm{abl}}^{(m)}
        -
        u_{i,\mathrm{orig}}^{(m)}
    \right]_{+}.
\end{align}
Here, $[x]_{+}=\max(x,0)$ and the bars denote averages over available
variants. Thus, $r_{i,u}^{(m)}$ measures the increase in method-$m$
uncertainty after truncating user history, while $r_{i,v}^{(m)}$ measures the
increase after ablating item content. 

Methods that provide explicit finite user-history and item-content
\texttt{source\_scores} use those values directly. The operational
dominant-control label $d_i$ and the method prediction $\hat{d}_i^{(m)}$ are
defined in Eqs.~(\ref{eq:dominant-control}) and~
(\ref{eq:predicted-control}), respectively. Equal method responses produce a
tie prediction, and cases with tied clipped NDCG@10 differences are excluded
from the dominant-control evaluation. The interpretation is limited to
agreement within the user-history and item-content controls.

\subsection{Coverage and Health Checks}
\label{sec:appendix-coverage}

Every evaluated Llama UQ artifact contains all 45,163 generated ranking rows
with 100\% join coverage. Restricting the popularity-tail condition to the
1,312 eligible tail-target records leaves 42,475 records in the final
evaluation: 4,000 Warm-start, 20,000 Sparse-history, 17,163 Item-text
robustness, and 1,312 tail-target records. Each Qwen replication artifact
contains the same 4,000 Warm-start rows. 

All evaluated artifacts report zero unmatched \texttt{record\_id}s and zero
schema errors. Base records use unique identifiers. Malformed generated rankings are repaired deterministically: 337 Llama records are completed by appending missing labels, and two Qwen Warm-start outputs fall back to the original candidate order after a failed retry.

\subsection{Supplementary Robustness Checks}
\label{sec:appendix-robustness-checks}

\paragraph{Matched Error@K/TopKMass.}
Table~\ref{tab:appendix-topk-alignment} compares Top5Mass, Top10Mass, and
Top15Mass against Error@5, Error@10, and Error@15 on the shared 4,000 Llama
Warm-start records. Within the TopKMass family, the cutoff-aligned score is
highest for each monitored Error@K event. Relative to the strongest
mismatched cutoff, the aligned score improves AUROC by $0.022$ for Error@5,
$0.012$ for Error@10, and $0.036$ for Error@15.

\begin{table}[t]
\centering
\small
\setlength{\tabcolsep}{7pt}
\begin{tabular}{lrrr}
\toprule
Score & Error@5 & Error@10 & Error@15 \\
\midrule
Top5Mass & \textbf{0.656} & 0.609 & 0.583 \\
Top10Mass & 0.634 & \textbf{0.628} & 0.592 \\
Top15Mass & 0.624 & 0.616 & \textbf{0.628} \\
\bottomrule
\end{tabular}
\caption{\textbf{Warm-start Error@K/TopKMass alignment.}
Error-AUROC on the shared 4,000 Llama Warm-start records.}
\label{tab:appendix-topk-alignment}
\end{table}

\paragraph{Instance-Clustered Confidence Intervals.}
Sparse-history and Item-text robustness contain multiple variants derived from
the same base instances. Their bootstrap resampling therefore samples
\texttt{dataset}$\times$\texttt{instance\_id} clusters, paired across methods.
The tail-target subset contains one retained record per instance.
Table~\ref{tab:appendix-cluster-bootstrap} reports the 200-resample confidence
intervals used in the condition-level analysis.

\begin{table*}[t]
\centering
\small
\setlength{\tabcolsep}{7pt}
\renewcommand{\arraystretch}{1.08}
\begin{tabular}{lrrll}
\toprule
Condition
& Records
& Clusters
& \shortstack{Top10Mass AUROC\\{[95\% CI]}}
& \shortstack{Best alternative\\{[95\% CI]}} \\
\midrule
Sparse-history
& 20,000 & 4,000
& 0.634 [0.621, 0.646]
& Decomp.: 0.561 [0.550, 0.570] \\

Tail-target subset
& 1,312 & 1,312
& 0.547 [0.520, 0.580]
& Perturb.: 0.527 [0.497, 0.556] \\

Item-text robustness
& 17,163 & 4,000
& 0.640 [0.626, 0.653]
& Decomp.: 0.551 [0.540, 0.562] \\
\bottomrule
\end{tabular}
\caption{\textbf{Instance-clustered paired bootstrap intervals.}
Aggregate Llama Error-AUROC with 95\% confidence intervals for the
non-reference evaluation conditions.}
\label{tab:appendix-cluster-bootstrap}
\end{table*}

\paragraph{NDCG@10-Difference Sensitivity.}
Table~\ref{tab:appendix-source-gap} restricts the paired-control evaluation to
cases whose clipped user-history and item-content NDCG@10 differences differ by
more than the stated threshold. As the minimum gap increases, Perturbation
Rank Instability retains nearly complete directional coverage while its
dominant-control accuracy, Source-AUROC, and rank correlation increase.

\begin{table}[t]
\centering
\small
\setlength{\tabcolsep}{5pt}
\begin{tabular}{rrrrrr}
\toprule
Gap & $n$ & DirCov & D-Acc & Source-AUROC & $\rho$ \\
\midrule
0.00 & 1,819 & 0.987 & 0.709 & 0.675 & 0.456 \\
0.05 & 1,522 & 0.988 & 0.739 & 0.701 & 0.477 \\
0.10 & 1,124 & 0.989 & 0.775 & 0.739 & 0.498 \\
\bottomrule
\end{tabular}
\caption{\textbf{Perturbation response under NDCG@10-difference gap thresholds.}}
\label{tab:appendix-source-gap}
\end{table}

At a threshold of $0.10$, Source-AUROC increases by $0.064$,
dominant-control accuracy by $0.066$, and rank correlation by $0.042$ relative
to the full non-tied set; directional coverage changes by only $0.002$. This
sensitivity analysis remains scoped to the two constructed controls. Because clipping can set a negative NDCG@10 difference to zero, we additionally
restrict the analysis to the 863 cases in which both user-history and item-content NDCG@10 differences are positive before clipping. Perturbation Rank
Instability remains the strongest method, with dominant-control accuracy of
$0.714$ and Source-AUROC of $0.671$, compared with $0.709$ and $0.675$ on
the full non-tied set. This positive-positive comparison covers the eight UQ signals in
Table~\ref{tab:source-response}.

\begin{figure*}[t]
\centering
\includegraphics[width=0.92\textwidth]
{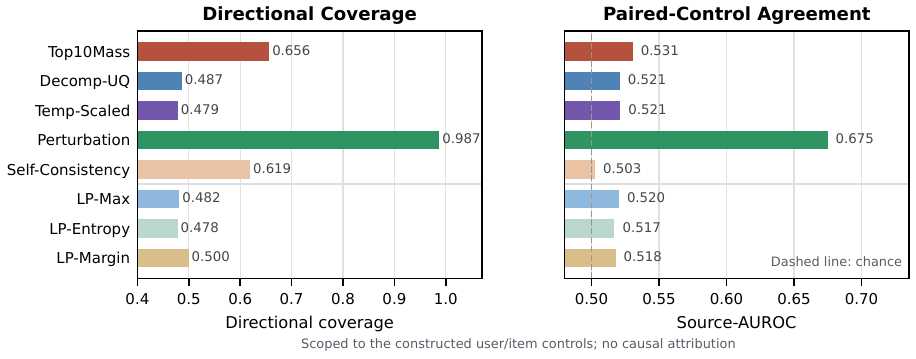}
\caption{\textbf{Paired-control response agreement on
Llama-3.1-8B-Instruct.}
Directional coverage and Source-AUROC for the user-history and item-content
controls.}
\label{fig:paired-control-response}
\end{figure*}

\subsection{Sanity Checks}
\label{sec:appendix-sanity}

We include three non-deployable references to verify the orientation and
implementation of the Error-AUROC evaluator. These checks are not included in
the benchmark ranking.

\begin{table}[t]
\centering
\small
\setlength{\tabcolsep}{4pt}
\begin{tabular}{lrr}
\toprule
Reference & Error-AUROC & Coverage \\
\midrule
Random score & 0.495 & 1.000 \\
Parse-failure indicator & 0.501 & 1.000 \\
Oracle Error@10 & 1.000 & 1.000 \\
\bottomrule
\end{tabular}
\caption{\textbf{Sanity checks for Error-AUROC.}
Random and parse-failure references remain near chance; the oracle reaches the
expected upper bound.}
\label{tab:appendix-sanity}
\end{table}

All three references cover the 42,475 evaluation-eligible Llama records.
Random scoring and the parse-failure indicator remain near chance, while the
oracle confirms that larger uncertainty is aligned with the Error@10 event.

\paragraph{Below-Chance AUROC Diagnostics.}
All UQ scores retain their predefined orientation, with larger values
indicating greater nominal uncertainty. On Qwen Warm-start, several methods
with Error-AUROC below $0.5$ also assign lower mean uncertainty to Error@10
cases than to correct cases. For example, the mean uncertainty is $0.0847$
versus $0.1943$ for Top10Mass and $0.1849$ versus $0.2707$ for
Self-consistency. In contrast, Decomposed LLM UQ, with an Error-AUROC of
$0.559$, has mean uncertainty of $0.5192$ on Error@10 cases and $0.4917$ on
correct cases. This pattern is consistent with the below-chance AUROC values
reflecting an inverse association between the nominal uncertainty signal and
Error@10 in these settings. For both backbones, the one-letter scores use next-token logits for
labels A--T and supported bare and space-prefixed token forms, followed
by normalization over the 20 labels
(Appendix~\ref{sec:appendix-uq-definitions}).
The Qwen probe places its maximum probability on label A in 81.9\% of records.
Relative to the one-letter Top10Mass score on the issued set, using the
probe's own top-10 set increases Error-AUROC by 0.121.
Using first-slot probabilities reconstructed under the ranking prompt
increases it by 0.071
(Appendix~\ref{sec:appendix-same-inference}).
Averaging the probe distributions across candidate orders increases it by 0.190.
These comparisons show sensitivity to the monitored set, probability source,
and candidate order.

\subsection{Ranking-Aligned and Component Checks}
\label{sec:appendix-derived-checks}

We derive component-level diagnostics from the existing artifacts without
additional LLM inference. Top10Mass is included as the ranking-aligned
reference; the remaining rows isolate the base top-1 probability,
decomposition components, and self-consistency top-1 summaries.

\begin{table}[!t]
\centering
\small
\setlength{\tabcolsep}{4pt}
\begin{tabular}{lr}
\toprule
Check & Error-AUROC \\
\midrule
LabelProb-BaseTop1 & 0.584 \\
LabelProb-Top10Mass & 0.634 \\
Decomp. recommendation component & 0.542 \\
Decomp. prompt component & 0.483 \\
Decomp. prompt ratio & 0.478 \\
SC top-1 disagreement & 0.508 \\
SC top-1 entropy & 0.509 \\
SC top-1 FSD & 0.507 \\
\bottomrule
\end{tabular}
\caption{\textbf{Derived ranking-aligned and component checks.}
Error-AUROC pooled over all 42,475 evaluation-eligible Llama records.}
\label{tab:appendix-derived-checks}
\end{table}

In the pooled component evaluation, Top10Mass improves Error-AUROC from
$0.584$ for BaseTop1 to $0.634$, a gain of $0.050$. The recommendation
component of Decomposed LLM UQ reaches $0.542$, its prompt components reach
$0.478$--$0.483$, and the self-consistency top-1 components reach
$0.507$--$0.509$. These values contextualize the main scalar results without
adding new LLM inference.

\subsection{Released Artifacts and Data Card}
\label{sec:appendix-release-organization}
\label{sec:appendix-dataset-view}
\label{sec:appendix-matched-restoration}
\label{sec:appendix-extra-csvs}
\label{sec:appendix-release}

The anonymized artifact package is organized around fixed base predictions,
record-aligned UQ outputs, evaluation tables, diagnostic tables, manifests,
schemas, run metadata, and reconstruction notes. The complete
dataset-by-condition Error-AUROC breakdown and per-method paired-control results are provided as machine-readable CSV artifacts. The diagnostic files
also include paired-control confusion summaries, per-instance NDCG@10 differences, sanity checks, component checks, and coverage reports.

Following data-documentation and reproducibility practices
\citep{bender2018data,gebru2021datasheets,pineau2021improving,
lhoest2021datasets}, the anonymized artifact package includes evaluator scripts, schema documentation, manifests, run metadata, reconstruction instructions, and
data-card notes. The data card documents the source
datasets, evaluation conditions, record counts, intended use, and known
limitations of the fixed-ranking monitoring benchmark. Redistribution of raw
item text follows the licenses of MIND, Amazon, and Yelp. When raw text cannot
be redistributed, the release will provide identifiers, hashes, rankings,
scores, and reconstruction scripts where permitted.

\subsection{\texorpdfstring{Signed NDCG@10 Differences and Ranking-Response Baselines}{Signed NDCG@10 Differences and Ranking-Response Baselines}}
\label{sec:appendix-signed}

The signed analysis supplements the clipped evaluation in
Eqs.~(\ref{eq:user-restoration-gain})--(\ref{eq:predicted-control}) and the
positive-positive analysis in Appendix~\ref{sec:appendix-robustness-checks}.
Its user-history and item-content NDCG@10 differences are
$N_{i,\mathrm{full}}-\overline{N}_{i,\mathrm{trunc}}$ and
$N_{i,\mathrm{orig}}-\overline{N}_{i,\mathrm{abl}}$, respectively.
The signed label selects the larger difference.
Equal differences are excluded using the evaluator's tie convention.
Negative differences indicate that mean NDCG@10 across the controlled
variants exceeds NDCG@10 under the original input.

For the signed evaluation, scalar responses are
$\bar{u}_{i,\mathrm{trunc}}^{(m)}-u_{i,\mathrm{full}}^{(m)}$ and
$\bar{u}_{i,\mathrm{abl}}^{(m)}-u_{i,\mathrm{orig}}^{(m)}$. Source-specific ranking
distances use the same definitions in both evaluations. Equal response scores follow the
tie-prediction convention in Appendix~\ref{sec:appendix-source-response}. The
majority-class reference predicts the more frequent source label within each evaluation
set.

\begin{figure}[t]
\centering
\includegraphics[width=0.85\textwidth]{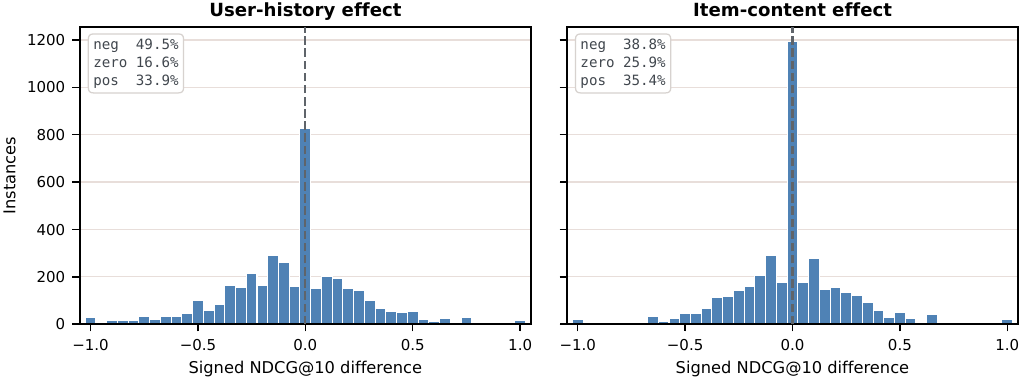}
\caption{\textbf{Distributions of signed NDCG@10 differences on Llama.}
Negative differences indicate higher mean NDCG@10 across the controlled
variants relative to the original input.}
\label{fig:appendix-signed-effects}
\end{figure}

Table~\ref{tab:appendix-signed-effects}(a) reports effect signs, and panel (b)
reports ties and dominant-source label counts. Figure~\ref{fig:appendix-signed-effects}
shows the signed-effect distributions. Clipping preserves the user/item label of every
case that remains non-tied, across all datasets and both backbones. Its effect on
evaluation membership is the creation of additional ties. The signed Llama labels are
more frequently item-content dominant.

\begin{table}[t]
\centering
\footnotesize
\setlength{\tabcolsep}{6pt}
\renewcommand{\arraystretch}{1.12}
\begin{tabular}{lrcc}
\toprule
\multicolumn{4}{l}{\textit{(a) Effect signs: positive / negative / zero (\%)}}\\
Dataset & $n$ & User history & Item content \\
\midrule
MIND-20 & 1,000 & 39.9/44.8/15.3 & 34.7/33.2/32.1 \\
Books & 1,000 & 27.7/57.9/14.4 & 39.0/45.2/15.8 \\
Movies & 1,000 & 39.0/47.4/13.6 & 30.7/34.4/34.9 \\
Yelp & 1,000 & 29.1/47.7/23.2 & 37.1/42.2/20.7 \\
\midrule
Llama, all & 4,000 & 33.9/49.5/16.6 & 35.4/38.8/25.9 \\
Qwen, all & 4,000 & 28.8/40.0/31.2 & 28.1/34.6/37.2 \\
\bottomrule
\end{tabular}

\par\medskip
\begin{tabular}{lrrrr}
\toprule
\multicolumn{5}{l}{\textit{(b) Ties and dominant-source label counts}}\\
Dataset & Exact ties & Added by clipping & \shortstack{Clipped labels\\(user/item)} & \shortstack{Signed labels\\(user/item)} \\
\midrule
MIND-20 & 97 & 429 & 289/185 & 435/468 \\
Books & 87 & 461 & 187/265 & 318/595 \\
Movies & 66 & 464 & 247/223 & 440/494 \\
Yelp & 133 & 444 & 183/240 & 343/524 \\
\midrule
Llama, all & 383 & 1,798 & 906/913 & 1,536/2,081 \\
Qwen, all & 918 & 1,582 & 830/670 & 1,415/1,667 \\
\bottomrule
\end{tabular}
\caption{\textbf{Signed NDCG@10 differences and the impact of clipping.} Dataset rows
use Llama; the last row gives pooled Qwen results. Effect-sign percentages use all cases
in each row and are rounded to one decimal. Label counts exclude ties under the
corresponding label definition.}
\label{tab:appendix-signed-effects}
\end{table}

The non-UQ baselines apply top-5 Jaccard, normalized Kendall, or RBO distance to
the same original--controlled ranking pairs. Top-5 Jaccard uses one minus the top-5
intersection-to-union ratio. Normalized Kendall is the fraction of discordant item pairs
in the full ranking. RBO distance uses one minus the weighted prefix-overlap fraction,
with weights $p^{d-1}$ at depth $d$, $p=0.9$, and normalization by the sum of those
weights over the 20 ranking positions. Each source response averages its pairwise
distances over that source's available controlled variants. Top-10 Jaccard yields
exactly the same source scores as Perturbation Rank Instability on both backbones.

Table~\ref{tab:appendix-signed-dataset}(a) reports dataset-specific signed results
on Llama; panel (b) reports clipped Source-AUROC on both backbones. The Llama clipped
intervals use dataset-stratified bootstrap resampling. For Llama, the 95\% confidence
interval for the signed Source-AUROC of Perturbation Rank Instability is [0.484, 0.509].
The ranking distances show strong clipped agreement and near-chance signed source
discrimination. The dataset-specific PRI and RBO results follow the latter pattern.
Decomposed LLM UQ's D-Acc stays close to the majority-class reference across these
datasets. These comparisons characterize ranking sensitivity to the two controlled
evidence sources.

\begin{table}[t]
\centering
\footnotesize
\setlength{\tabcolsep}{4.5pt}
\renewcommand{\arraystretch}{1.12}
\begin{tabular}{lrccccc}
\toprule
\multicolumn{7}{l}{\textit{(a) Signed-label results by dataset on Llama}}\\
Dataset & $n$ & \shortstack{Majority\\D-Acc} & \shortstack{Decomp.\\D-Acc} & \shortstack{Decomp.\\Src-AUROC} & \shortstack{PRI\\Src-AUROC} & \shortstack{RBO\\Src-AUROC} \\
\midrule
MIND-20 & 903 & 0.518 & 0.525 & 0.520 & 0.487 & 0.507 \\
Books & 913 & 0.652 & 0.650 & 0.561 & 0.497 & 0.503 \\
Movies & 934 & 0.529 & 0.531 & 0.526 & 0.494 & 0.494 \\
Yelp & 867 & 0.604 & 0.615 & 0.558 & 0.492 & 0.500 \\
\bottomrule
\end{tabular}

\par\medskip
\begin{tabular}{lcc}
\toprule
\multicolumn{3}{l}{\textit{(b) Clipped Source-AUROC and the Qwen check}}\\
Signal & Llama ($n=1,819$) & Qwen ($n=1,500$) \\
\midrule
Perturbation Rank Instability & 0.675 [0.661, 0.689] & 0.634 \\
Kendall distance & 0.692 [0.678, 0.707] & 0.639 \\
RBO distance & 0.692 [0.677, 0.706] & 0.674 \\
\bottomrule
\end{tabular}
\caption{\textbf{Supplementary paired-control results.} Decomp. denotes Decomposed LLM
UQ with unclipped scalar responses in panel (a); PRI denotes Perturbation Rank
Instability. Brackets in panel (b) give 95\% confidence intervals. The Qwen diagnostic
uses auxiliary controlled rankings for the same Warm-start base instances.}
\label{tab:appendix-signed-dataset}
\end{table}

\subsection{\texorpdfstring{Evidence Removal and Robustness-Setting Details}{Evidence Removal and Robustness-Setting Details}}
\label{sec:appendix-removal}

Positional outputs follow the candidate order or its reverse. They account for
46.1\% of Llama Warm-start rankings (7.1\% identity and 39.0\% reverse) and 48.9\%
of Qwen Warm-start rankings (all identity). Their per-dataset Error@10 on Llama is
0.46--0.51. Table~\ref{tab:appendix-removal}(a) reports the reduction in positional
share after restricting Books history to the most recent interaction and after
removing Yelp review summaries on Qwen. On the pooled non-positional Llama Warm-start subset, Top10Mass reaches an Error-AUROC of 0.612. This subset excludes rankings that follow the candidate order or its reverse.

Table~\ref{tab:appendix-removal}(b) reports NDCG@10 among non-positional outputs
in each evaluated condition. History truncation improves this conditional quality on
Books and Yelp; MIND-20 and Movies show no corresponding gain. These comparisons evaluate shorter histories that retain the most recent interactions. Item-text gains are field-specific, as illustrated by removing Books author information
and Yelp review summaries.

Table~\ref{tab:appendix-removal}(c) reports original-input NDCG@10 separately for
Movies instances eligible for each field ablation and the remaining instances. The
eligible subsets already have higher original NDCG@10. Field availability therefore
contributes to the aggregate Item-text result through variant composition. These
diagnostics supplement the condition-level results in Table~\ref{tab:base-ranking}.

\begin{table}[t]
\centering
\footnotesize
\setlength{\tabcolsep}{4pt}
\renewcommand{\arraystretch}{1.12}

\begin{tabular*}{0.80\textwidth}{@{\extracolsep{\fill}}lllrr@{}}
\toprule
\multicolumn{5}{@{}l}{\textit{(a) Positional-output share (\%)}} \\
Backbone & Dataset & Controlled input & Original & Controlled \\
\midrule
Llama & Books & Most recent interaction & 29.6 & 7.1 \\
Qwen & Yelp & Review summary removed & 46.9 & 10.1 \\
\bottomrule
\end{tabular*}

\par\medskip

\begin{tabular*}{0.80\textwidth}{@{\extracolsep{\fill}}llrr@{}}
\toprule
\multicolumn{4}{@{}l}{\textit{(b) NDCG@10 among non-positional outputs on Llama}} \\
Dataset & Controlled input & Original & Controlled \\
\midrule
Books & History truncated & 0.508 & 0.556--0.599 \\
Yelp & History truncated & 0.465 & 0.488--0.505 \\
Books & Author removed & 0.508 & 0.547 \\
Yelp & Review summary removed & 0.465 & 0.570 \\
\bottomrule
\end{tabular*}

\par\medskip

\begin{tabular*}{0.80\textwidth}{@{\extracolsep{\fill}}lrrr@{}}
\toprule
\multicolumn{4}{@{}l}{\textit{(c) Field availability and original-input quality on Llama Movies}} \\
& & \multicolumn{2}{c@{}}{Original NDCG@10} \\
\cmidrule(l){3-4}
Field & Eligible instances & Eligible & Remaining \\
\midrule
Title & 587 & 0.381 & 0.224 \\
Description & 576 & 0.384 & 0.223 \\
\bottomrule
\end{tabular*}

\caption{\textbf{Evidence-removal diagnostics.} Original uses full history and original
item text; Controlled uses the input change specified in the row. Panel (b) conditions
on the non-positional outputs of each condition; ranges span the four history-truncation
variants. Panel (c) compares original-input quality across field-availability subsets.}
\label{tab:appendix-removal}
\end{table}

For the settings in Section~\ref{sec:method-datasets}, Sparse-history includes the
full-history record and the four truncation variants, giving five records per instance
and 20,000 records overall. Item-text variants require the removed field to be present
for at least one candidate. The Movies title and description coverage in
Table~\ref{tab:appendix-removal}(c) follows this rule. Tail cutoffs are computed over items
with training interactions, and eligibility includes targets with zero training
interactions. The MIND-20/Books/Movies/Yelp tail counts are 870/221/129/92; 856 of the
870 MIND-20 targets have zero training interactions.


\subsection{\texorpdfstring{Same-Prompt Listwise Scores}{Same-Prompt Listwise Scores}}
\label{sec:appendix-same-inference}

We reconstruct probabilities under the original listwise ranking prompt for the
same 4,000 Warm-start records on each backbone. Prompt hashes match the base run. The
stored output is teacher-forced, and we read valid-label distributions at emitted label
slots using the same bare and space-prefixed token forms as LabelProb, together with
emitted-token log-probabilities. The issued ranking remains fixed throughout.

First-slot top-10 mass is one minus the first-slot probability mass on the issued
top-10 labels. First-slot max probability uses one minus the maximum valid-label
probability. Sequence NLL@10 sums the negative log-probabilities of the first ten
emitted labels. The probe-own-set control applies the top-10-mass calculation to the ten
highest-probability labels of the separate one-letter distribution.
Table~\ref{tab:appendix-same-inference} reports these scores using the Warm-start
evaluator and 1,000 dataset-stratified paired bootstrap resamples.

\begin{table}[t]
\centering
\footnotesize
\setlength{\tabcolsep}{4pt}
\renewcommand{\arraystretch}{1.12}
\begin{tabular}{p{0.23\textwidth}p{0.27\textwidth}cc}
\toprule
Score & Probability source / set & Llama & Qwen \\
\midrule
LabelProb-Top10Mass & One-letter prompt / issued top-10 & 0.628 [0.611, 0.644] & 0.447 [0.430, 0.464] \\
Probe own top-10 mass & One-letter prompt / probe top-10 & 0.525 [0.508, 0.544] & 0.568 [0.550, 0.585] \\
First-slot top-10 mass & Ranking prompt / issued top-10 & 0.606 [0.589, 0.624] & 0.518 [0.500, 0.535] \\
First-slot max probability & Ranking prompt & 0.527 [0.509, 0.545] & 0.486 [0.469, 0.505] \\
Sequence NLL@10 & Ranking prompt & 0.399 [0.381, 0.416] & 0.306 [0.290, 0.322] \\
\bottomrule
\end{tabular}
\caption{\textbf{Same-prompt and cross-elicitation scores.} Warm-start Error-AUROC with
95\% confidence intervals. Ranking-prompt scores use teacher-forced reconstructions of
the stored output.}
\label{tab:appendix-same-inference}
\end{table}

The two comparisons have backbone-dependent directions. On Qwen, using the issued
set lowers Error-AUROC by 0.121 relative to the probe's own set; reconstructing
first-slot top-10 mass under the ranking prompt raises it by 0.071 relative to
cross-elicitation Top10Mass. Sequence NLL@10 is below chance on both backbones under its
predefined orientation.

These probabilities are teacher-forced reconstructions of the stored outputs.
Greedy next-token agreement is 94.4\% for Llama and 99.5\% for Qwen. Using the original
transformers 5.7.0 version with KV-cache decoding, exact greedy regeneration matches 19
of 50 checked Llama outputs.


\subsection{\texorpdfstring{Self-Consistency with the Base Ranking}{Self-Consistency with the Base Ranking}}
\label{sec:appendix-sc-base}

We use the existing five sampled rankings to evaluate the two base-inclusive
variants described after Eq.~(\ref{eq:sc}). Base-anchored disagreement averages the
top-10 Jaccard distance between the issued ranking and each sample. Samples plus base
applies the original pairwise calculation to those six rankings.
Table~\ref{tab:appendix-sc-base} reports the original and both base-inclusive scores.

\begin{table}[t]
\centering
\footnotesize
\setlength{\tabcolsep}{4pt}
\renewcommand{\arraystretch}{1.12}

\begin{tabular}{lcccc}
\toprule
\multicolumn{5}{l}{\textit{(a) Llama Error-AUROC}} \\
Variant & Warm-start & Sparse-history & Tail-target & Item-text \\
\midrule
Samples only
& 0.533 [0.515, 0.552] & 0.536 & 0.506 & 0.523 \\
Base-anchored
& 0.555 [0.538, 0.573] & 0.552 & 0.524 & 0.554 \\
Samples plus base
& 0.553 [0.536, 0.571] & 0.555 & 0.523 & 0.550 \\
\bottomrule
\end{tabular}

\par\medskip

\begin{tabular}{lc}
\toprule
\multicolumn{2}{l}{\textit{(b) Qwen Warm-start Error-AUROC}} \\
Variant & Warm-start \\
\midrule
Samples only & 0.381 [0.362, 0.399] \\
Base-anchored & 0.365 [0.347, 0.383] \\
Samples plus base & 0.376 [0.356, 0.394] \\
\bottomrule
\end{tabular}

\caption{\textbf{Self-consistency with the base ranking.}
Error-AUROC; Warm-start entries include 95\% confidence intervals.
Qwen is evaluated on Warm-start.}
\label{tab:appendix-sc-base}
\end{table}

Relative to samples-only self-consistency, the two variants increase Warm-start
Error-AUROC by 0.022 and 0.020 on Llama and decrease it by 0.016 and 0.005 on Qwen,
respectively. Both variants remain below Top10Mass in each Llama condition and below
Decomposed LLM UQ on Qwen Warm-start.


\subsection{\texorpdfstring{Inference Cost}{Inference Cost}}
\label{sec:appendix-cost}

Table~\ref{tab:appendix-cost} separates additional model calls from the
base-ranking call. The four LabelProb signals share a forward pass; temperature scaling
reuses its output and fits a one-dimensional temperature using labeled calibration data.
Decomposed LLM UQ uses candidate-order forward passes, and self-consistency uses sampled
generations. Perturbation Rank Instability reuses the benchmark's controlled Llama
rankings. Its controlled-generation requirement is 7 calls for MIND-20, 8 for Books and
Yelp, and 5--7 for Movies.

\begin{table}[htbp]
\centering
\footnotesize
\setlength{\tabcolsep}{3.8pt}
\renewcommand{\arraystretch}{1.12}
\begin{tabular}{p{0.25\textwidth}p{0.20\textwidth}p{0.22\textwidth}rr}
\toprule
Signal & \shortstack{Additional calls\\per record} & Type & \shortstack{Llama\\(45,163)} & \shortstack{Qwen\\(4,000)} \\
\midrule
LabelProb family & 1 (shared) & Forward pass & 8h02m & 1h28m \\
Temperature-scaled LabelProb & 0 (reused output) & Offline fit & -- & -- \\
Decomposed LLM UQ & 5 & Forward passes & 22h11m & 3h26m \\
Self-consistency & 5 & Sampled generations & 121h53m & 13h25m \\
Perturbation Rank Instability & 7.3 on average & Controlled greedy generations & Reused & 17h43m$^\dagger$ \\
\midrule
Base ranking & 1 base call (reference) & Greedy generation & 24h32m & 2h36m \\
\bottomrule
\end{tabular}
\caption{\textbf{Inference requirements and logged wall-clock totals.} Times cover the
indicated record sets. $^\dagger$The Qwen perturbation time covers 41,163 controlled
rankings.}
\label{tab:appendix-cost}
\end{table}

All runs use a single A100 80GB. Wall-clock totals are indicative run costs under
method-specific batch settings. The Llama timing set includes generated records before
tail-target eligibility filtering; the Qwen controlled-ranking total is identified
separately in Table~\ref{tab:appendix-cost}.

\end{document}